\documentclass[twocolumn]{aastex61}

\newcommand{\Wms}{W~m$^{-2}$}
\newcommand{\Porb}{\ensuremath{P_{\mathrm{orb}}}}
\newcommand{\Prot}{\ensuremath{P_{\mathrm{rot}}}}
\newcommand{\pss}{\ensuremath{\phi_{\mathrm{ss}}}}
\newcommand{\pobs}{\ensuremath{\phi_{\mathrm{obs}}}}

\shorttitle{Obliquities of Warm Jupiters}
\shortauthors{Rauscher}

\begin{document}

\title{Models of Warm Jupiter Atmospheres: Observable Signatures of Obliquity}

\email{erausche@umich.edu}

\author[0000-0003-3963-9672]{Emily Rauscher}
\affil{Department of Astronomy, University of Michigan, 1085
  S. University Ave., Ann Arbor, MI 48109, USA}
  
\begin{abstract}

We present three-dimensional atmospheric circulation models of a
hypothetical ``warm Jupiter'' planet, for a range of possible
obliquities from 0-90\degr. We model a Jupiter-mass planet on a 10-day
orbit around a Sun-like star, since this hypothetical planet
sits at the boundary between planets for which we expect that tidal
forces should have aligned their rotation axes with their orbital axes
(i.e., ones with zero obliquity) and planets whose timescale for tidal
alignment is longer than the typical age of an exoplanet system. 
In line with observational progress, which is pushing atmospheric
characterization to planets on longer orbital periods, we calculate
the observable signatures of obliquity for a transiting
warm Jupiter: in orbital phase curves of thermal emission and in the
hemispheric flux gradients that could be measured 
by eclipse mapping. For both of these predicted measurements, the
signal that we would see depends strongly on our viewing geometry
relative to the orientation of the planet's rotation axis, and we
thoroughly identify the degeneracies that result.
We compare these signals to the predicted
sensitivities of current and future instruments and determine that the
James Webb Space Telescope should be able to constrain the obliquities
of nearby warm Jupiters to be small (if $\le 10\degr$) or to directly
measure them if significantly non-zero ($\ge 30\degr$), using
the technique of eclipse mapping.  For a bright target and
  assuming photon-limited precision, this could be done with a
single secondary eclipse observation.

\end{abstract}


\section{Introduction}

Exoplanet atmospheric characterization began with a hot Jupiter
\citep{Charbonneau2002} and the majority of all measurements so far
have been of the atmospheres of these Jupiter-mass planets that orbit
less than 0.1 AU from their host stars. In addition to being
inherently interesting, hot Jupiters are the biggest, brightest
transiting planets and so the easiest to detect and
characterize. However, we are entering an era where 
transit searches are finding increasing numbers of cooler planets on
longer orbital periods and, importantly, ones around bright enough
stars that we can extend atmospheric characterization beyond hot
Jupiters. This is also facilitated by a better understanding of how to best
use current instruments for atmospheric measurements, as well as the
much-anticipated launch of new facilities, especially the
\textit{James Webb Space Telescope}. As a field were are poised to
expand our understanding of exoplanets to the increasingly diverse
possiblities we should expect as we study planets on longer orbital periods.

As of this writing, there are almost three dozen warm Jupiters known
to transit their host star,\footnote{NASA Exoplanet
Archive, as of 03/20/2017 ({\tt
  https://exoplanetarchive.ipac.caltech.edu/})} where we define ``warm
Jupiter'' somewhat  
arbitrarily to mean a planet with a radius greater than 0.5 Jupiter
radii\footnote{Uranus and Neptune have radii of $\sim 0.35\
  R_{\mathrm{Jupiter}}$.} and a zero-albedo equilibrium temperature
between 500-1000 K. These planets are all around stars bright
enough for us to estimate their effective temperatures (necessary to
estimate the planets' equilibrium temperatures). The current brightest
planet of this population is WASP-69b, a 
1 Jupiter-radius, 0.3 Jupiter-mass planet with an equilibrium
temperature of $\sim$960 K around a $V=9.87$ magnitude star
\citep{Anderson2014}. The next two brightest, with V magnitudes
between 10.5-11 are HAT-P-17b (1 $R_J$, 0.5 $M_J$, $T_{eq} \sim 790$
K, and $e=0.34$; Howard et al. 2012) and WASP-84b (0.9 $R_J$, 0.7
$M_J$, and $T_{eq} \sim 800$ K; Anderson et al. 2014). We expect that
this population of known warm Jupiters should grow as current transit
searches continue and new ones begin. In particular, missions such as
the Transiting Exoplanet Survey Satellite (TESS) and the PLAnetary
Transits and Oscillations of stars (PLATO) satellite should identify
more planets that transit bright stars, improving our ability to
characterize these worlds.

A significant difference between the populations of ``warm'' and
``hot'' Jupiters is that the overwhelmingly strong tidal forces
experienced by hot Jupiters will no longer be so overwhelming for the
longer orbital period, warm population. As a consequence, whereas we
generally assume that the rotation rates of hot Jupiters are synchronized with
their orbital rates and their rotation axes are aligned with
their orbital axes, these assumptions no longer hold for warm Jupiters
and we should expect a range of possible values for these
parameters. One observable expression of this
decreasing influence of tidal forces on longer period planets is the
increase in their range of orbital eccentricities \citep[see review
by][]{Winn2015}. As we expand 
observational and theoretical efforts out from hot Jupiters there are
three planetary properties that were fixed for the hot Jupiter
population, but that we should expect to create additional diversity
among the longer period ``warm Jupiters'': eccentricity, rotation
rate, and obliquity.\footnote{Here we use the term obliquity to refer
  to the angle of the planet's rotation axis away from its orbital
  axis. This is different from the widely studied (and more easily
  observable) obliquity between a star's rotation axis and the
  exoplanet's orbital axis \citep[see review by][]{Winn2015}.} There has
been previous work studying different eccentricities
\citep{Kataria2013} and rotation rates
\citep{Showman2015} for warm Jupiters; here we focus on non-zero obliquities.

Previous work studying non-zero obliquities of
exoplanets has generally fallen into one of two main categories. The
first includes papers that study the influence of obliquity on
atmospheric circulation patterns, often with an emphasis on Earth-like
planet and their habitability
\citep[e.g.,][]{Williams1996,Williams2003,Spiegel2009,Ferreira2014,Linsenmeier2015,Shields2016}. 
The second category contains papers that identify means by which we may be able to measure or
constrain the obliquities of exoplanets, either using a presumed
atmospheric state or analytic models that allow for any
general spatial pattern. \citet{Seager2002}
demonstrated that the oblateness of a planet, due to rotation, can
impart a small signal on its transit light curve, and the obliquity of
the planet influences the shape and strength of this feature. This
method was subsequently used to observationally constrain the oblateness
of several exoplanets and a brown dwarf
companion \citep{Carter2010,Zhu2014}. There have also been studies of ways
to constrain a transiting or directly imaged planet's obliquity with future
instrumentation, including signatures of obliquity within scattered
light curves \citep{Kawahara2010,Kawahara2016}, polarimetric
observations \citep{deKok2011}, and the planet's Rossiter-McLaughlin
effect during secondary eclipse \citep{Nikolov2015}. 

In a more general
approach, \citet{Cowan2013a} and 
\citet{Schwartz2016} presented useful analytic frameworks 
for identifying how the intrinsic properties of exoplanets (e.g., their
rotation rates, inclinations, obliquities, albedos, etc.) couple with
the extrinsic viewing orientation (related to the planet's orbital
parameters) in order to shape observed orbital and rotational
photometric variations, for emitted or reflected light. They
explicitly identified how the spatial information on the planet is
expressed in the temporal information of the light curve, as well as
which spatial patterns are necessarily (i.e., mathematically) invisible in these light curves.

\citet{Gaidos2004} is a rare work that fits within both categories of
obliquity research; they simulated the influence of non-zero obliquity
(as well as non-zero eccentricity) on a planet's atmospheric
structure, using energy balance models (considering a range of thermal
inertias), and also predicted observable consequences, calculating thermal phase
curves as a function of viewing orientation. While these authors
focused on terrestrial planets with Earth-like temperatures, some of
their results regarding the influence of viewing orientation on
observed seasonality are reproduced in our predictions, as we will
describe below. They also found significant degeneracy between the
effects of obliquity, thermal inertia, and viewing orientation; we
uniquely identify secondary eclipse maps as a means by which we may be
able to break these degeneracies. In addition, our atmospheric model
is more complex and the thermal inertia is not an input but a prediction (albeit
influenced by other, more basic, planetary parameters we
choose). \citet{Cowan2012b} expanded on this by using two
detailed three-dimensional circulation models for the Earth (one
similar to current-day conditions and one representative of a snowball
state) to simulate orbital thermal phase curves, as viewed from
different orientations. While they did not vary obliquity or
eccentricity, their models did include self-consistent heat transport
and so could test the influence of thermal inertia by comparing their
temperate and snowball models.

In this work we identify the next set of observable
planets whose obliquities may be measurable (``warm Jupiters'') and
present a joint analysis of their atmospheric state and observability
(both via thermal phase curves and eclipse maps).
We start our exploration of non-zero obliquities by modeling a
hypothetical Jupiter-like planet on a 10-day orbit around a 
Sun-like star. We choose this orbital period because it corresponds to
where the timescales for tidal circularization and axial alignment
\citep[which are of similar magnitude, e.g.][]{Peale1999}
may be comparable to the typical lifetime of an exoplanet
system.  Starting with the timescale for tidal spin-down of a planet
  \citep{Guillot1996}, $\tau$, and solving for the planet's orbital
  period, we find:
\begin{equation}
P \sim \left( \frac{\tau}{Q} \frac{16 \pi^4 R_p^3}{G \omega_p M_p} \right)^{1/4}.
\end{equation}
This formulation disguises any complexity of the planet's tidal
dissipation by using a single parameter, $Q$, and we are also ignoring
any orbital changes that may happen over time, so we can use this to
estimate the orbital period at which the timescale for tidal effects
may be comparable to the age of the system (setting that equal to
$\tau$), but this is not a strict prediction for a specific boundary
in physical conditions.  Nevertheless, using the mass, radius, and
current spin of Jupiter (as an estimate for the initial spin
rate, $\omega_p$), an estimate for the tidal dissipation factor of
$Q\sim10^6$, and assuming an age on the order of Gyr, we
solve for an orbital period of $P \sim 10$ days.
This is then roughly the orbit at which we would not know
whether or not to expect that tidal forces should have aligned a planet's
obliquity, and so this presents an interesting potential target for
characterization. 

The equilibrium temperature for a planet on a 10-day orbit around a
Solar analog is $\sim$900 K. Interestingly, this is 
near the $\sim$1000 K boundary identified by \citet{Thorngren2016},
cooler than which we may be able to assess the metallicities of
exoplanets from their bulk densities, without the (still unknown) hot
Jupiter radius inflation mechanism confusing matters. In other words, the
planets that we may want to study in order to answer fundamental
questions about planet bulk compositions are also those planets for
which we should expect a range of non-zero obliquities.

In this paper we compare models for a hypothetical warm Jupiter at different
obliquities using a three-dimensional General Circulation Model (GCM).
In Section~\ref{sec:model} we discuss our 
modeling framework, including the treatment for non-zero obliquity.
In Section~\ref{sec:circ} we present the results from the GCM and
discuss the variation in circulation patterns between models.  We show
observable properties of each model in Section~\ref{sec:obs}: the
thermal phase curves (\ref{sec:pcurves}) and eclipse maps (\ref{sec:emaps}).  We also compare these
predicted signals to the sensitivities of current and future
instruments (\ref{sec:measurement}). Finally, in
Section~\ref{sec:conc} we summarize our results and the
potential for observationally constraining the obliquities of warm Jupiters.

\section{Atmospheric Circulation Model} \label{sec:model}

We present three-dimensional atmospheric circulation models for a
hypothetical planet with the mass, radius, and rotation rate of Jupiter, on a 10-day
period orbit around a Sun-like star.  We calculated the circulation patterns
using a numerical code that solves the primitive equations
of meteorology, with pressure as the vertical coordinate, and uses a
double-gray scheme for the radiative transfer, 
as described thoroughly in \citet{Rauscher2012b}. This code has been
updated to allow for the variable stellar irradiation patterns appropriate
for planets
with non-zero obliquity, as described in Section~\ref{sec:obl_flux}
below.  We ran models for obliquities from 0 to 90\degr, with all
other physical parameters remaining the same (as listed in
Table 1).  Specifically, we simulated planets with
obliquities of: 0\degr, 3\degr~(equal to Jupiter's obliquity),
10\degr, 30\degr~($\approx$ the obliquities of Earth, Saturn, and
Neptune), 60\degr, and 90\degr~(close to Uranus' obliquity).  We chose values for the
gray optical and infrared absorption coefficients such that the
analytic solution \citep{Guillot2010} roughly matches the
temperature-pressure profile for a hypothetical Jupiter-like planet at
0.1 AU, as calculated from a one-dimensional atmospheric model with detailed
radiative transfer \citep{Fortney2007}.

\begin{deluxetable}{lc}
\tablecaption{Model physical parameters}
\tablehead{
\colhead{Parameter}  &  \colhead{Value}
}
\startdata
Radius of the planet, $R_p$      & $6.986 \times 10^7$ m \\
Gravitational acceleration, $g$ & 26	                        m s$^{-2}$ \\
Orbital revolution rate, $\omega_{\mathrm{orb}}$ 		& $7.2722 \times 10^{-6}$ s$^{-1}$ \\
\ \ \ Corresponding period, \Porb & 10  day$_{\oplus}$ \\
Rotation rate, $\omega_{\mathrm{rot}}$ 		& $1.7587 \times 10^{-4}$ s$^{-1}$ \\
\ \ \ Corresponding period, \Prot & 10  hours \\
Incident flux at substellar point, $F_0$ & $1.36 \times 10^5$ W m$^{-2}$  \\
\ \ \ Corresponding temperature, $T_{\mathrm{irr}}$  & 1250 K \\
Internal heat flux, $F_{\uparrow \mathrm{IR}, \mathrm{int}}$ & 5.7 W m$^{-2}$ \\
\ \ \ Corresponding temperature, $T_{\mathrm{int}}$  &   100 K \\
Optical absorption coefficient, $\kappa_{\mathrm{vis}}$ & $2.6 \times 10^{-3}$ cm$^2$ g$^{-1}$ \\
\ \ \ Optical photosphere ($\tau_{\mathrm{vis}}=2/3$)                      & 667   mbar \\
Infrared absorption coefficient, $\kappa_{\mathrm{IR},0}$  & $5.2 \times 10^{-2}$  cm$^2$ g$^{-1}$ \\
\ \ \ Infrared photosphere ($\tau_{\mathrm{IR}}=2/3$)                      & 33   mbar \\
Specific gas constant, $\mathcal{R}$ &                                  3523    J kg$^{-1}$ K$^{-1}$ \\
$\mathcal{R}/c_P$ & 0.2857 \\
\enddata
\end{deluxetable}

Our standard model runs were performed at a horiztonal spectral
resolution of T42 (corresponding to $\sim$3\degr~resolution at the
equator) and with 30 vertical levels evenly sampled in $\log P$ from
100 bar to 1 mbar.  We applied numerical hyperdissipation as an
eighth-order operator on the wind and temperature fields \citep[as
  described in][]{Rauscher2012b}, with a dissipation timescale of 0.02
  \Prot, which we found to adequately remove noise at the smallest
  scales without overdamping the kinetic energy spectrum of the
  circulation.  We initialized the models using a horizontally
  uniform temperature profile, whose vertical (pressure) dependence
  was set to match the globally averaged analytic solution for our double-gray
  opacities \citep{Guillot2010}.  We started each run with the winds at rest and
  used 360 timesteps per \Prot~to simulate each planet for 3000
  orbital periods ($=72552$ \Prot), by which point the winds
  throughout the observable atmosphere had finished accelerating (for
  yearly averaged values).

We can calculate the expected scale of dynamical features in our
planet's global circulation using the Rossby deformation radius
\citep[e.g.,][]{SCM2010} and find that at the infrared photosphere
this scale is $\sim$8\degr.  While T42 resolution should be
sufficient to capture these features, we tested some limited models at
T63 and T85, corresponding to $\sim$2\degr~and
$\sim$1.4\degr~resolutions, respectively.  We found very similar
circulation patterns and temperature structures, without any smaller scale
detail emerging in these higher resolution runs, giving us confidence
that the T42 resolution is sufficient to model the global
  circulation pattern of this planet.

One unavoidable concern when modeling planetary
  atmospheres is whether there could be sub-grid physics, not captured
  within the numerical simulation, that could influence the global
  circulation of the planet.  While this may be of particular
  concern for hot Jupiters, where the intense stellar irradiation can
  drive super-sonic winds, potentially triggering sub-grid shocks and
  hydrodynamic instabilities \citep[e.g.,][]{Li2010,Fromang2016}, these
  particular effects  should be less important in the calmer
  atmospheric circulation of warm Jupiters.  (To preview our results:
  the maximum wind speeds we find are about 1 km s$^{-1}$, whereas
  the sound speeds in the atmosphere range from 2-3 km s$^{-1}$.)
  Nevertheless, we are fundamentally not resolving whatever form of
  sub-grid dissipation exists in these atmospheres, instead capturing
  it through our hyperdissipation parameter.  As there is no \emph{a
    priori} physical basis for calculating this parameter
  \citep{Cho1996}, our uncertainty in the correct value to use could,
  for example, result in errors in our maximum wind speeds
  \citep{Heng2011b}.  \citet{Mayne2014} also has a nice discussion of
  the implications of this and other assumptions used in hot Jupiter
  circulation modeling.  While issues of sub-grid physics are not
  unique to our models, and perhaps may be better behaved than in the
  hot Jupiter context, this is an important caveat that should be
  kept in mind.

\subsection{Stellar Flux Pattern for Non-zero Obliquities} \label{sec:obl_flux}

Our atmospheric circulation code has been used before to model the
atmospheres of planets with zero obliquity, for a variety of stellar
heating patterns.  In the case of hot Jupiters the stellar heating
pattern does not change with time (in the frame rotating
with the planet).  The flux has a maximum at the
substellar point, falls off as the cosine of the angle from the
substellar point, and is zero everywhere on the
planet's perpetual nightside.  In the case of non-synchronous
rotation \citep[as in][]{Rauscher2014} the same
hemispheric forcing pattern is used, but the longitude of the
substellar point varies with time.  If the planet is rotatingly
quickly enough that the time between consecutive star-rises\footnote{A
  ``star-rise'' is the exoplanet version of a sunrise.} is shorter than the radiative
timescale of the atmosphere, it is more appropriate to use a diurnally
averaged stellar heating pattern,\footnote{A diurnally averaged
  heating pattern is appropriate for most of the planets in the Solar
  System.} such that the equator receives more 
flux than the poles (for zero obliquity). This was the forcing pattern
we used for the models presented in \citet{May2016}.

Here we have updated our code to calculate the
appropriate stellar forcing patterns for planets with non-zero
obliquities.  For these planets the latitude of the substellar point
(\pss) has a yearly seasonal dependence given by:
\begin{equation}
\sin \pss = \sin \psi \sin (2\pi t/\Porb),  \label{eqn:pss}
\end{equation}
where $\psi$ is the planet's obliquity and the zero-point for time in
the orbit ($t=0$) is set to be the vernal 
equinox in the Northern hemisphere, such that the maximum Northern
excursion of the substellar latitude ($\pss=+\psi$) occurrs at the Northern
summer solstice ($t=0.25 \Porb$).  If the planet is
rotating quickly enough that it is appropriate to use a diurnally
averaged stellar flux pattern to heat the atmosphere, the downward
flux at the top of the atmosphere can be calculated as \citep[e.g.,][]{Liou}:
\begin{equation}
F(\phi) = \frac{F_o}{\pi} (\sin \phi \sin \pss \ H + \cos
\phi \cos \pss \sin H),
\end{equation}
where $F_0$ is the incident flux at the substellar point and $H$ is
the length of half a day (from star-rise to noon), measured 
in radians, for the given latitude: $H = \cos^{-1}(-\tan \phi \tan \pss)$.
The inverse cosine function is only defined when its argument is
within $[-1,1]$.  For latitudes greater/lesser than the substellar
latitude during Northern/Southern summer, $H=\pi$ and those regions of
the planet experience perpetual daylight for the season, while
similarly there are latitudes in the other hemisphere that will have $H=0$ and
experience seasonally perpetual nighttime.

Since we assume Jupiter's rotation rate for our hypothetical planet,
and a 10-day orbit around a Sun-like star, we can use the prescription
in \citet{Showman2015} to determine the appropriateness of using a
diurnally averaged stellar heating pattern in this case.  We do find
that the radiative timescale at the optical photosphere (the
$\tau=2/3$ surface for absorbed starlight)\footnote{More accurately,
\citet{Heng2014} showed that in the purely absorbing limit (as we
assume here) most of the starlight is absorbed at $\tau=0.63$.} is greater than the 
rotation period (favoring a diurnal average), but only by a factor of
$\sim$2. We tested the impact of using a diurnal stellar heating pattern
by performing limited versions of our zero and
90\degr~obliquity models in which we explicitly move the heating
pattern with time. In order to correctly
model the movement of the substellar point, it should not
advance more than one grid point in one timestep.  For the case of
zero obliquity, this is equivalent to requiring that the number of
timesteps per \Prot~is greater than the number of longitude points
(128 at our standard T42 resolution) times the number of timesteps between each time the
radiative transfer routine is calculated (10), times the
number of rotation periods in an orbital period minus one
($24.184-1$).  This minimum requirement forces us to use 30000
timesteps per rotation period, a factor of almost 100 greater than
(and so computationally slower than) our standard runs.  We found that
the atmospheric structure in these models was still mostly
axisymmetric, with wind and temperature patterns that look very
similar to those from the models with diurnally averaged heating patterns.
The deviations from axisymmetry were very small; for example, at the
infrared photosphere the temperatures in a single latitude circle are
always within 2\% of the average at that latitude. This supports the
use of a diurnal average for this planet's heating pattern.

\section{Comparison of circulation patterns} \label{sec:circ}

\begin{figure*}[ht!]
\begin{center}
\includegraphics[width=0.445\textwidth]{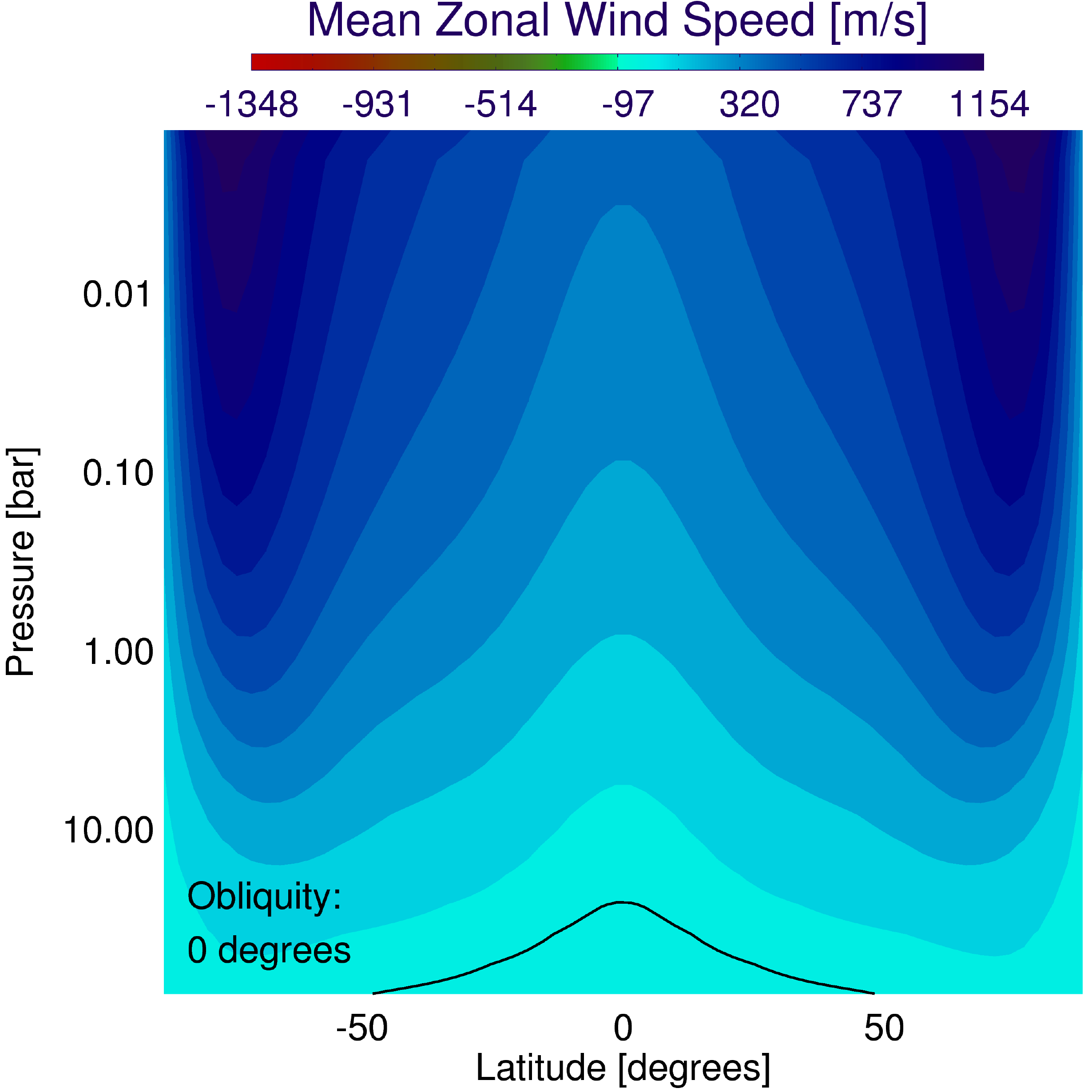}
\includegraphics[width=0.445\textwidth]{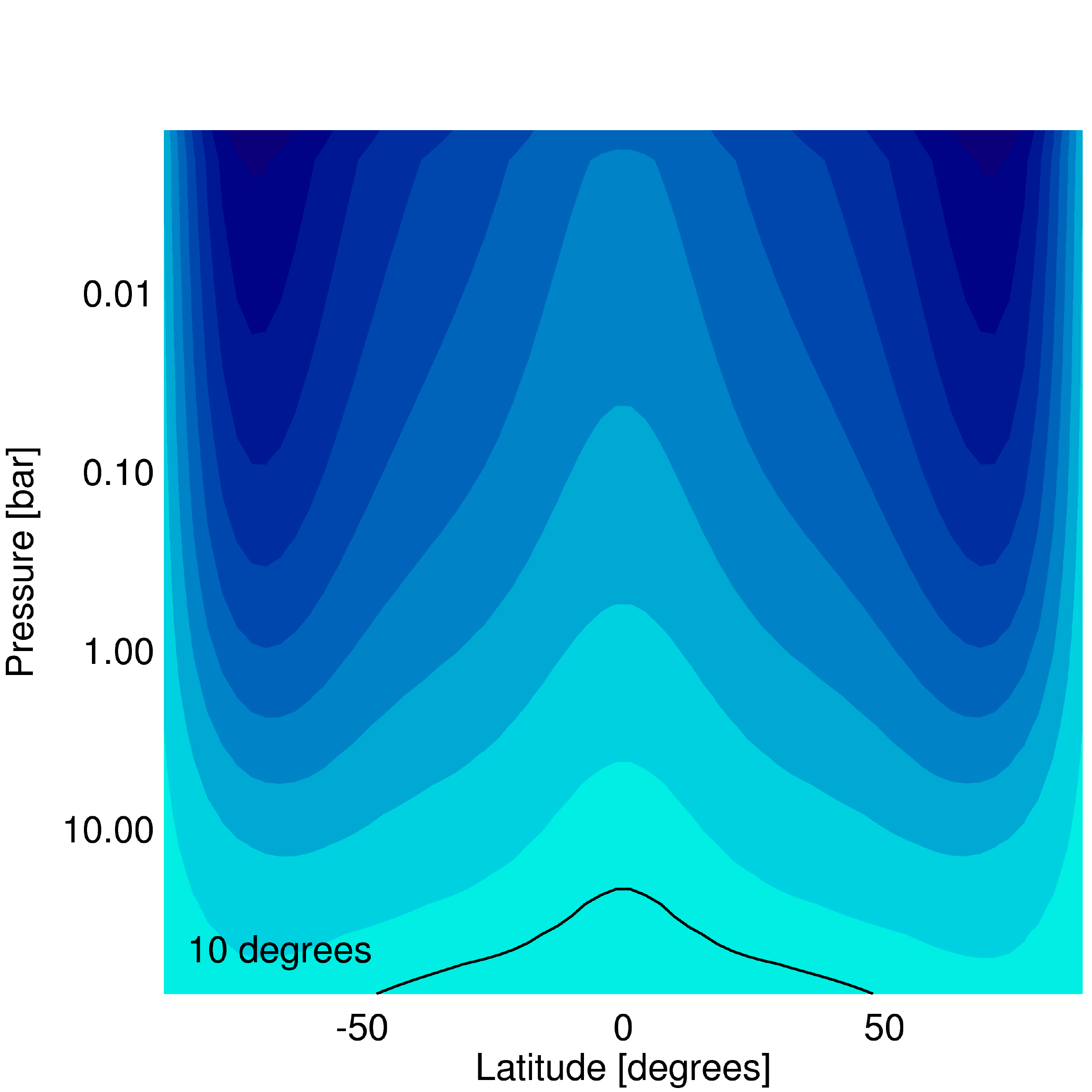} \\
\includegraphics[width=0.2514\textwidth]{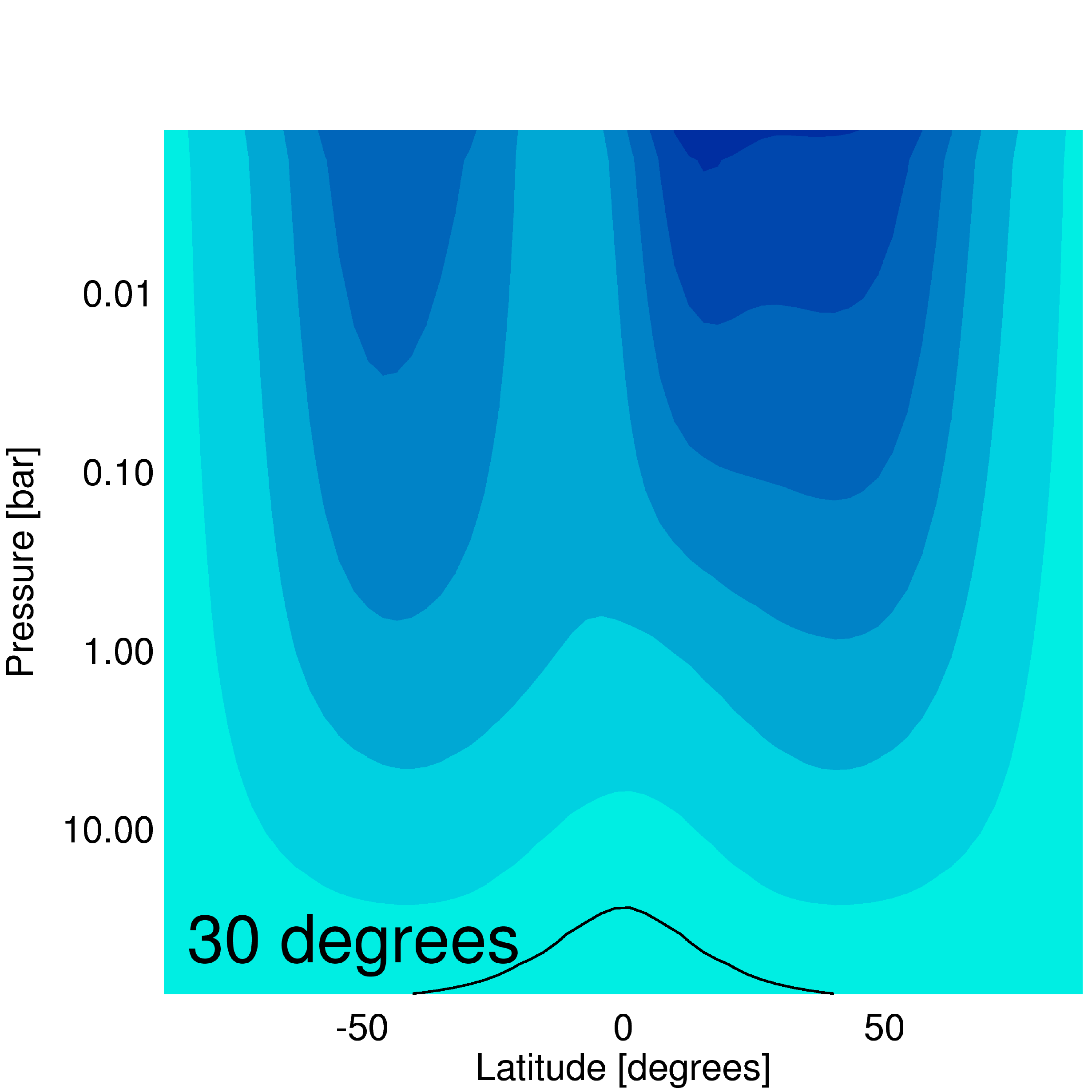}
\includegraphics[width=0.2155\textwidth]{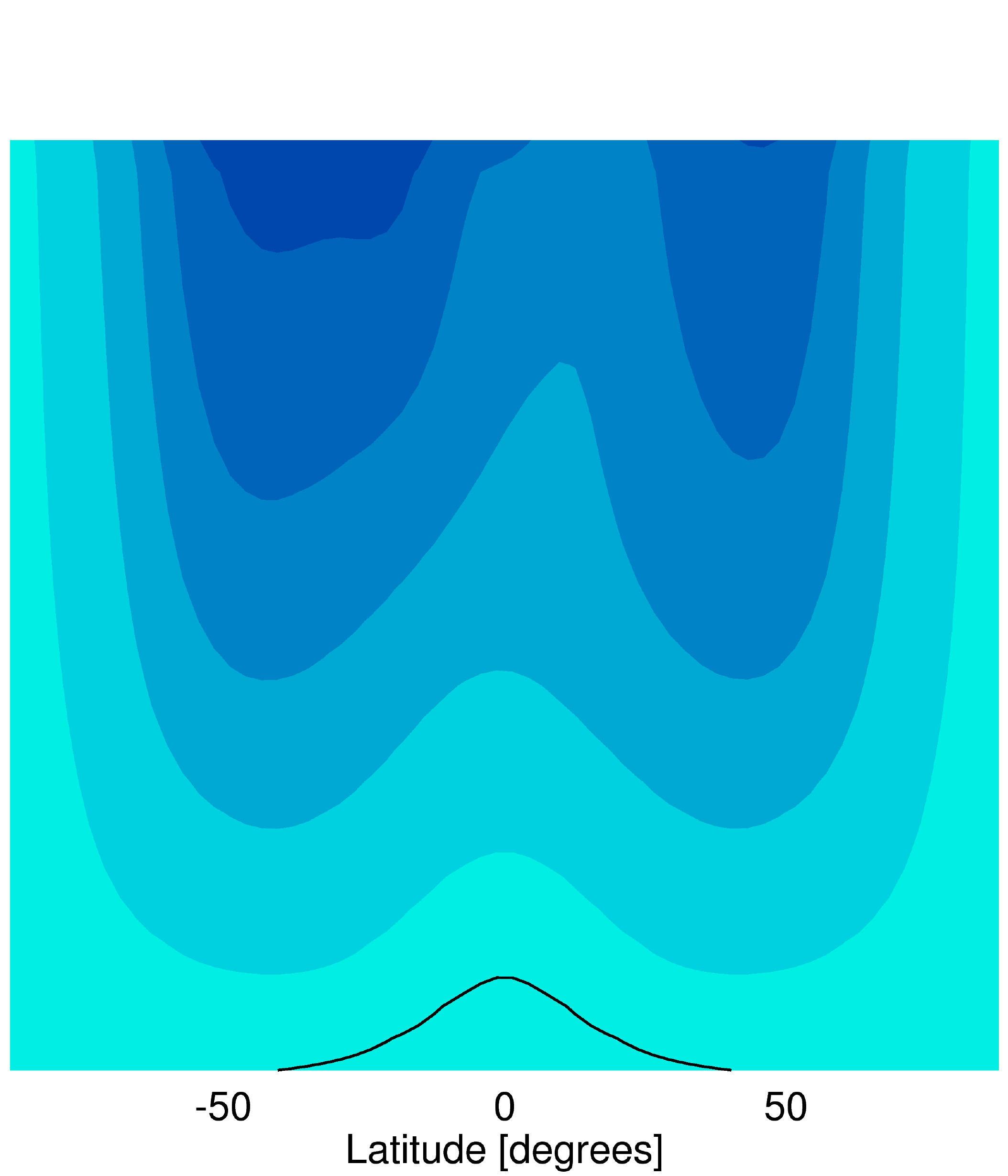}
\includegraphics[width=0.2155\textwidth]{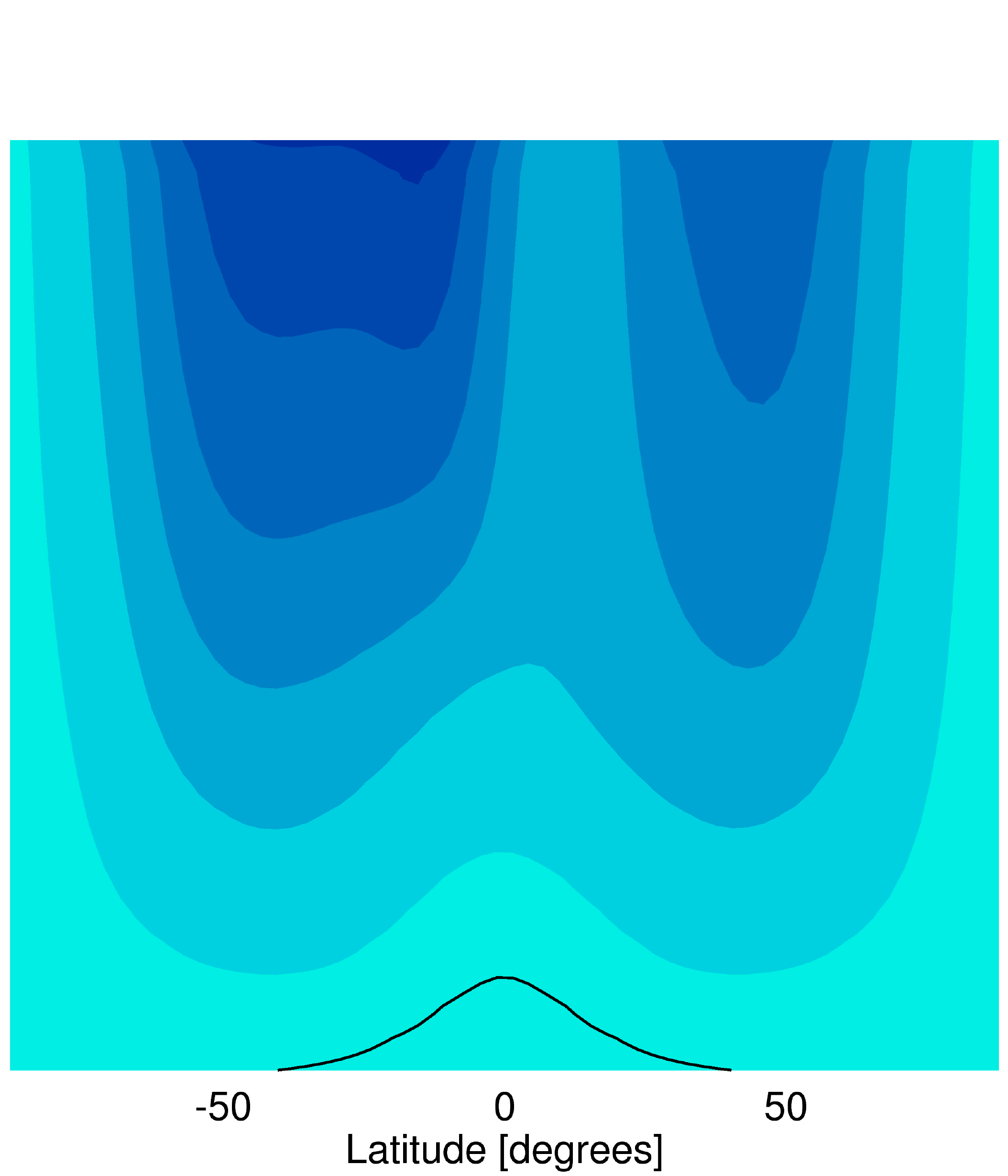}
\includegraphics[width=0.2155\textwidth]{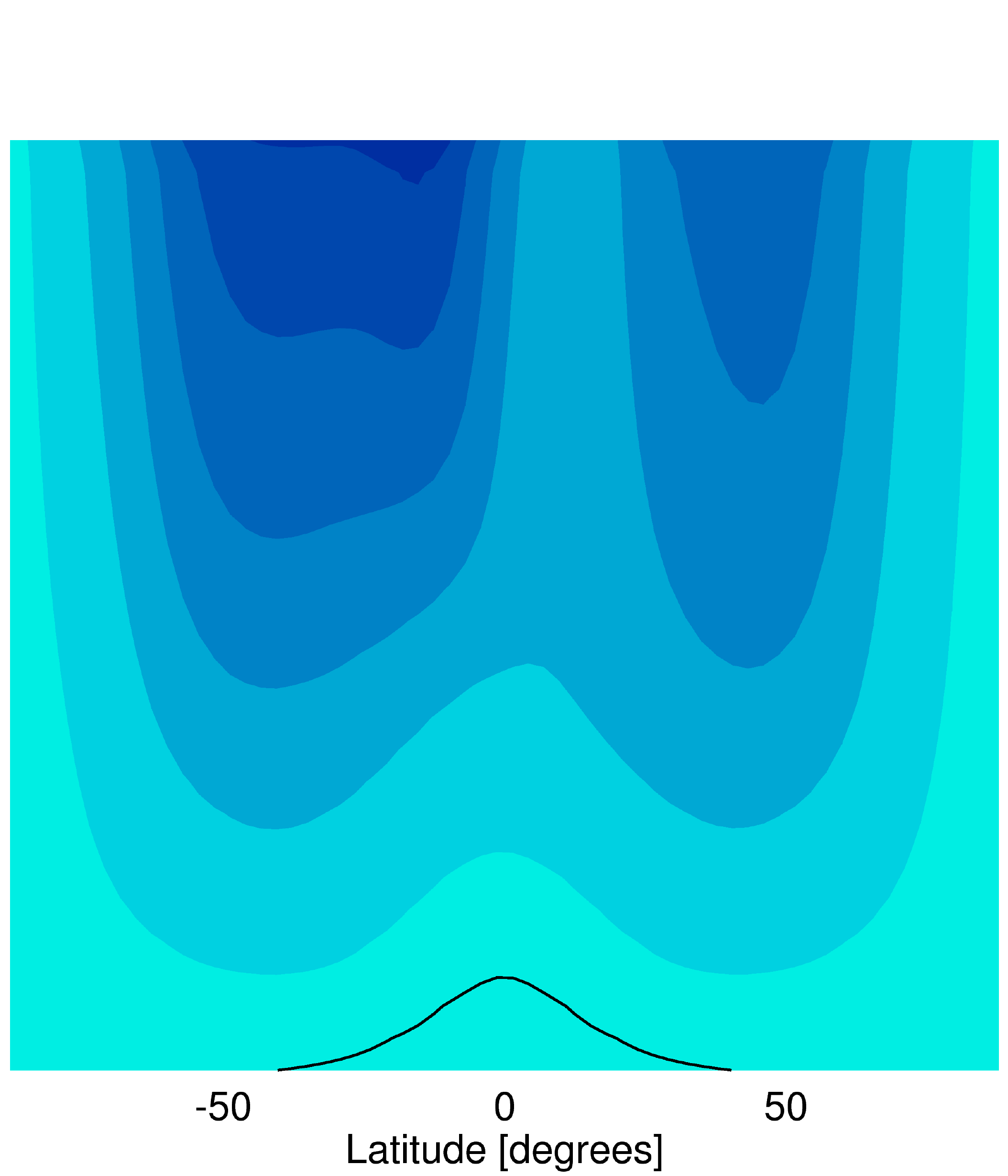} \\
\includegraphics[width=0.2514\textwidth]{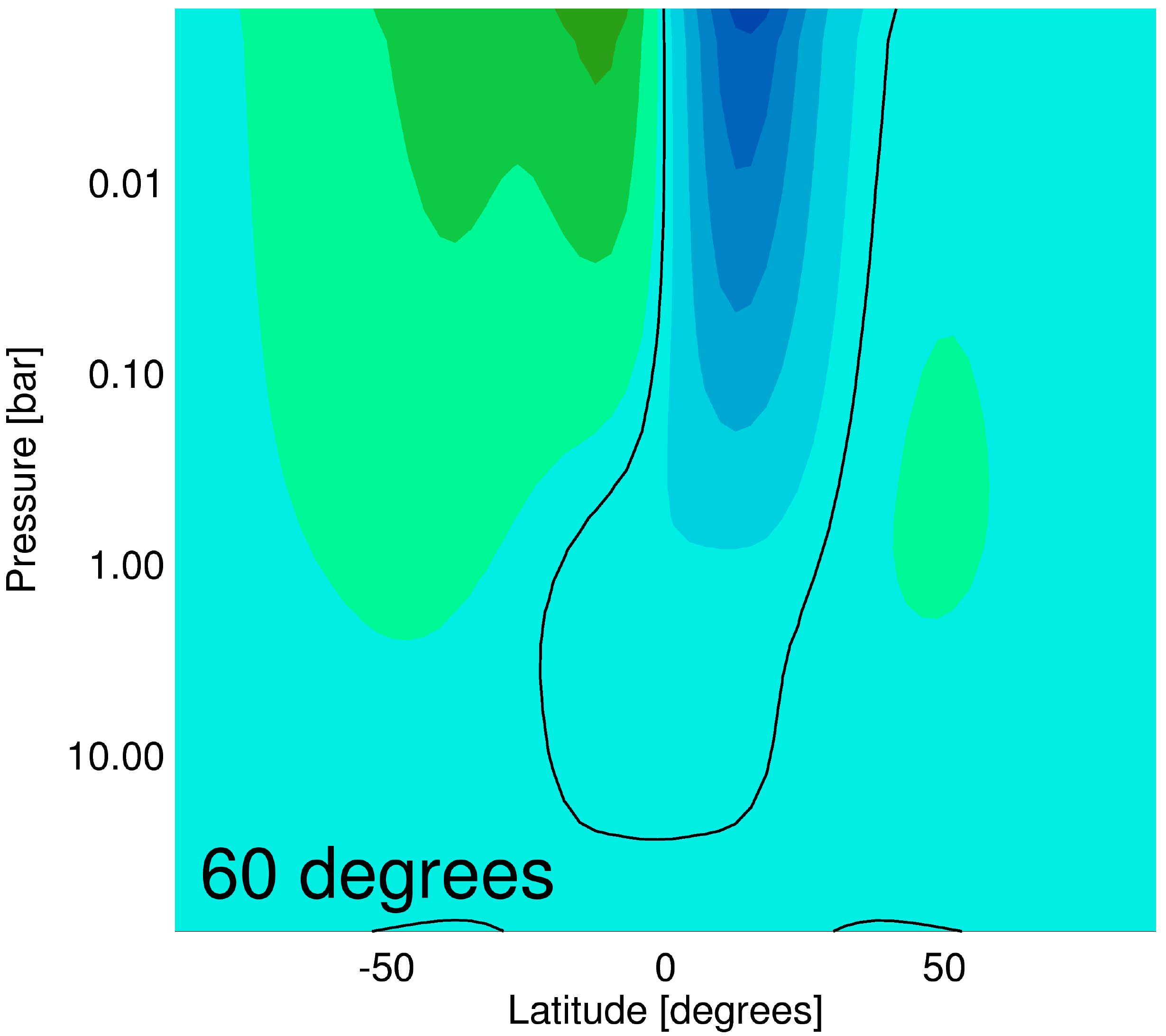}
\includegraphics[width=0.2155\textwidth]{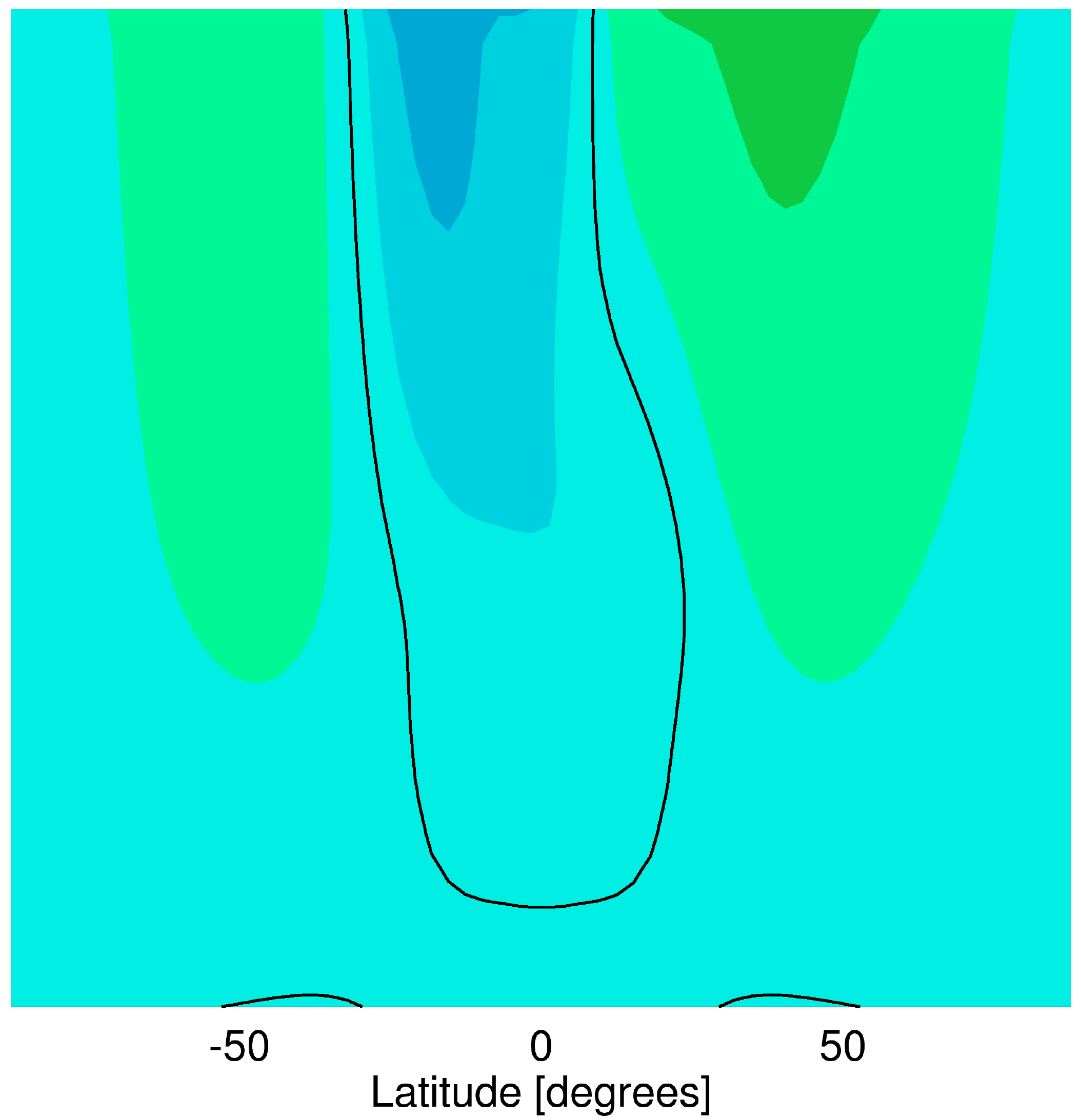}
\includegraphics[width=0.2155\textwidth]{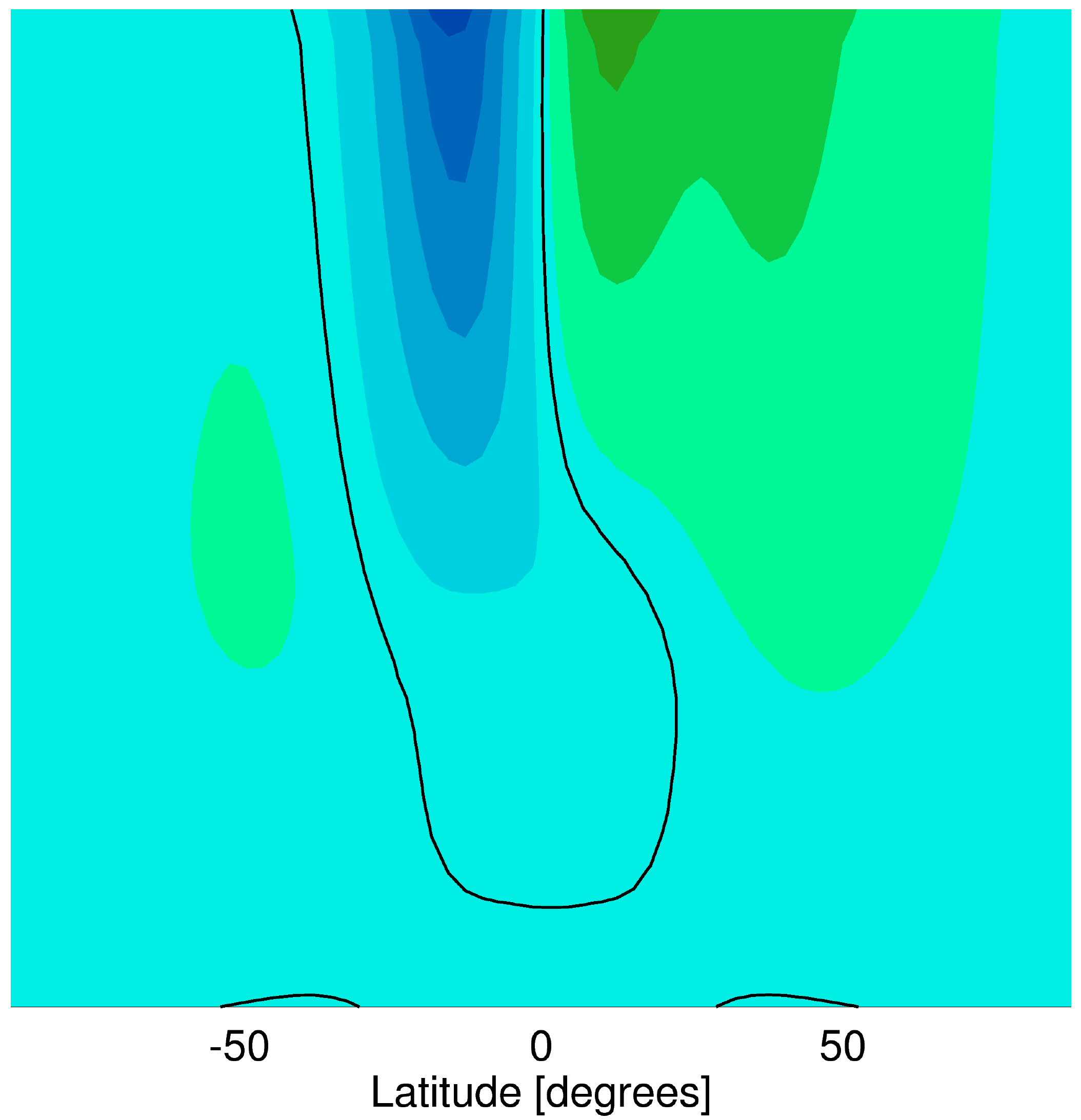}
\includegraphics[width=0.2155\textwidth]{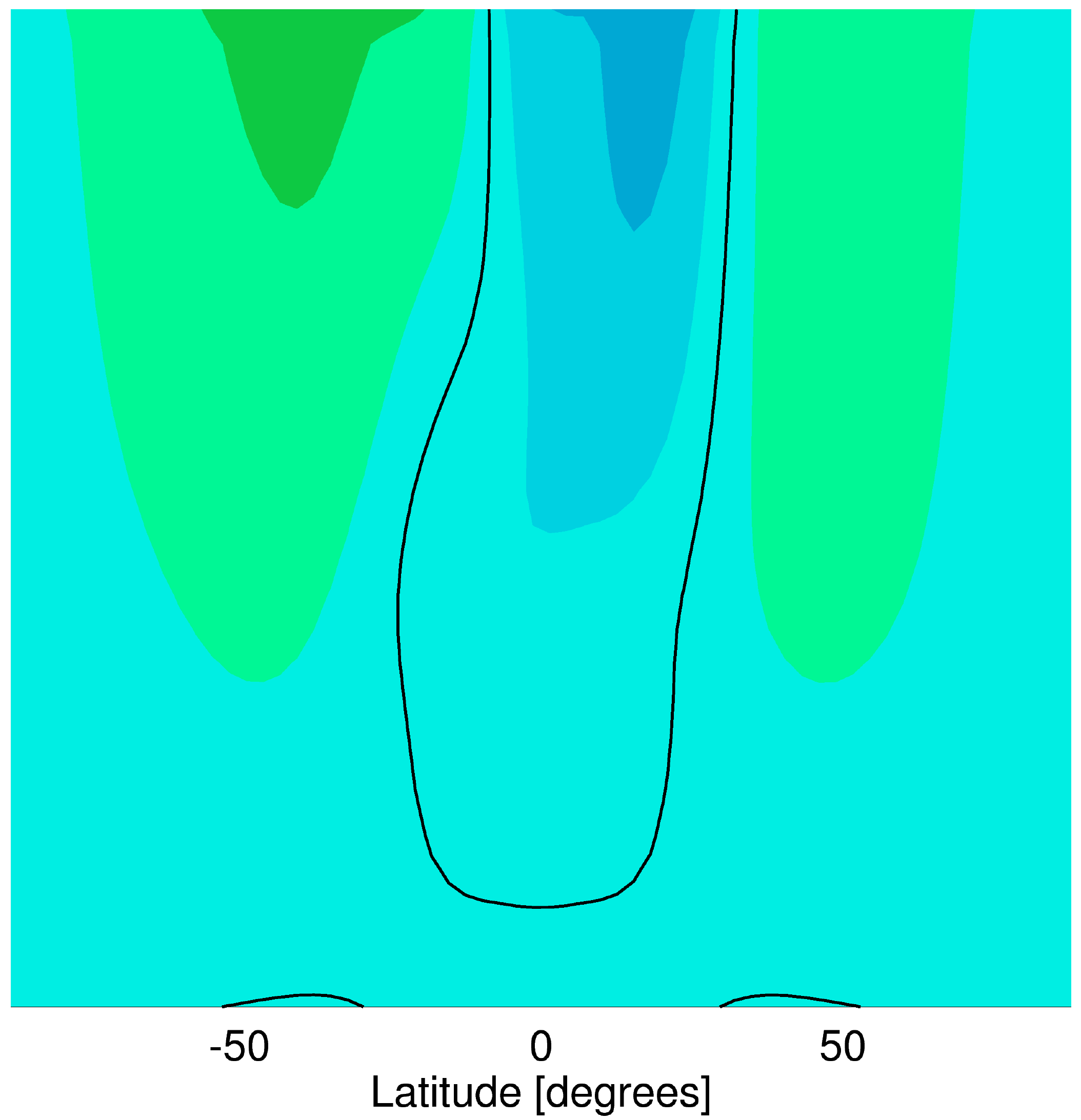} \\
\includegraphics[width=0.2514\textwidth]{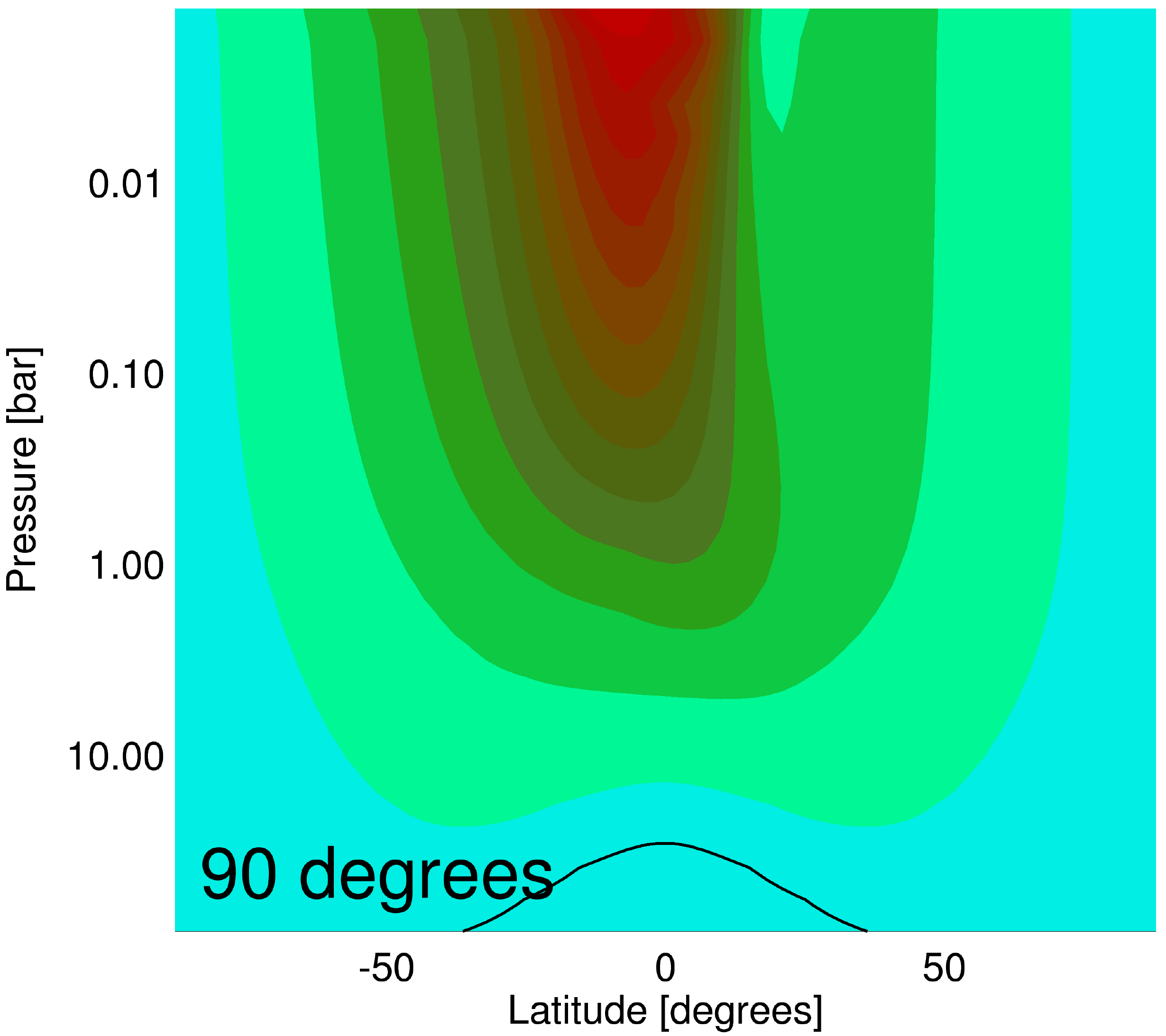}
\includegraphics[width=0.2155\textwidth]{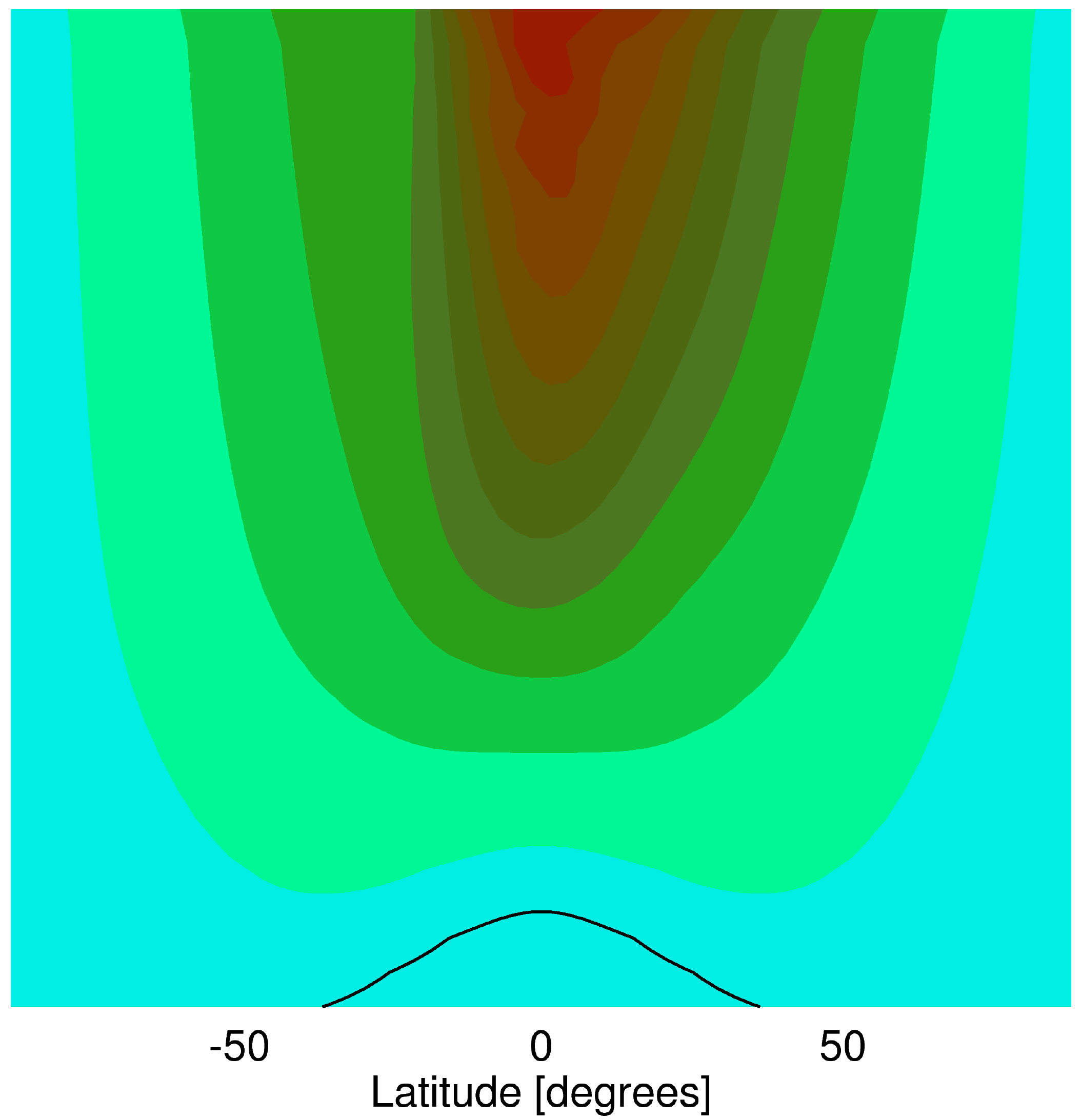}
\includegraphics[width=0.2155\textwidth]{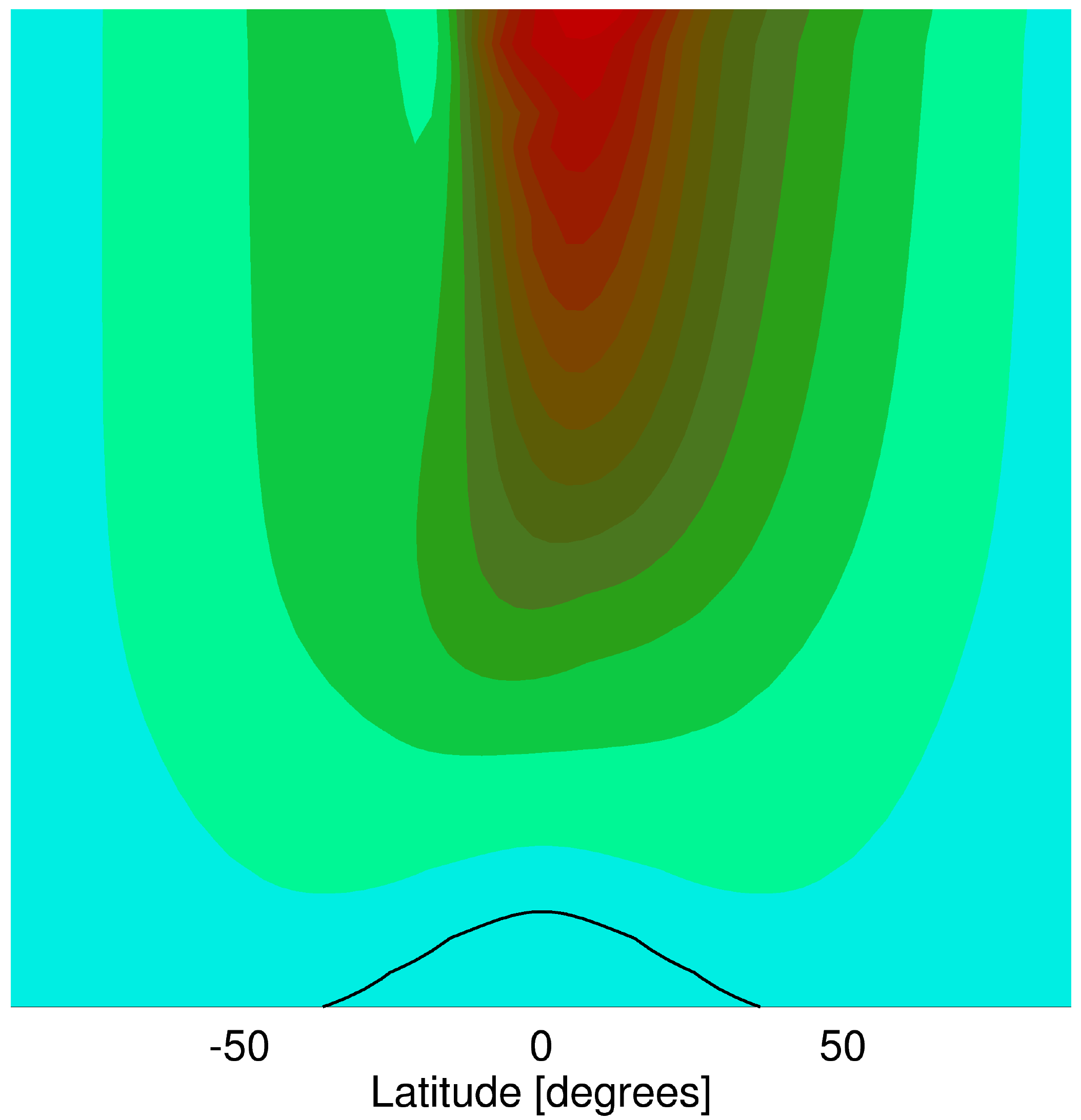}
\includegraphics[width=0.2155\textwidth]{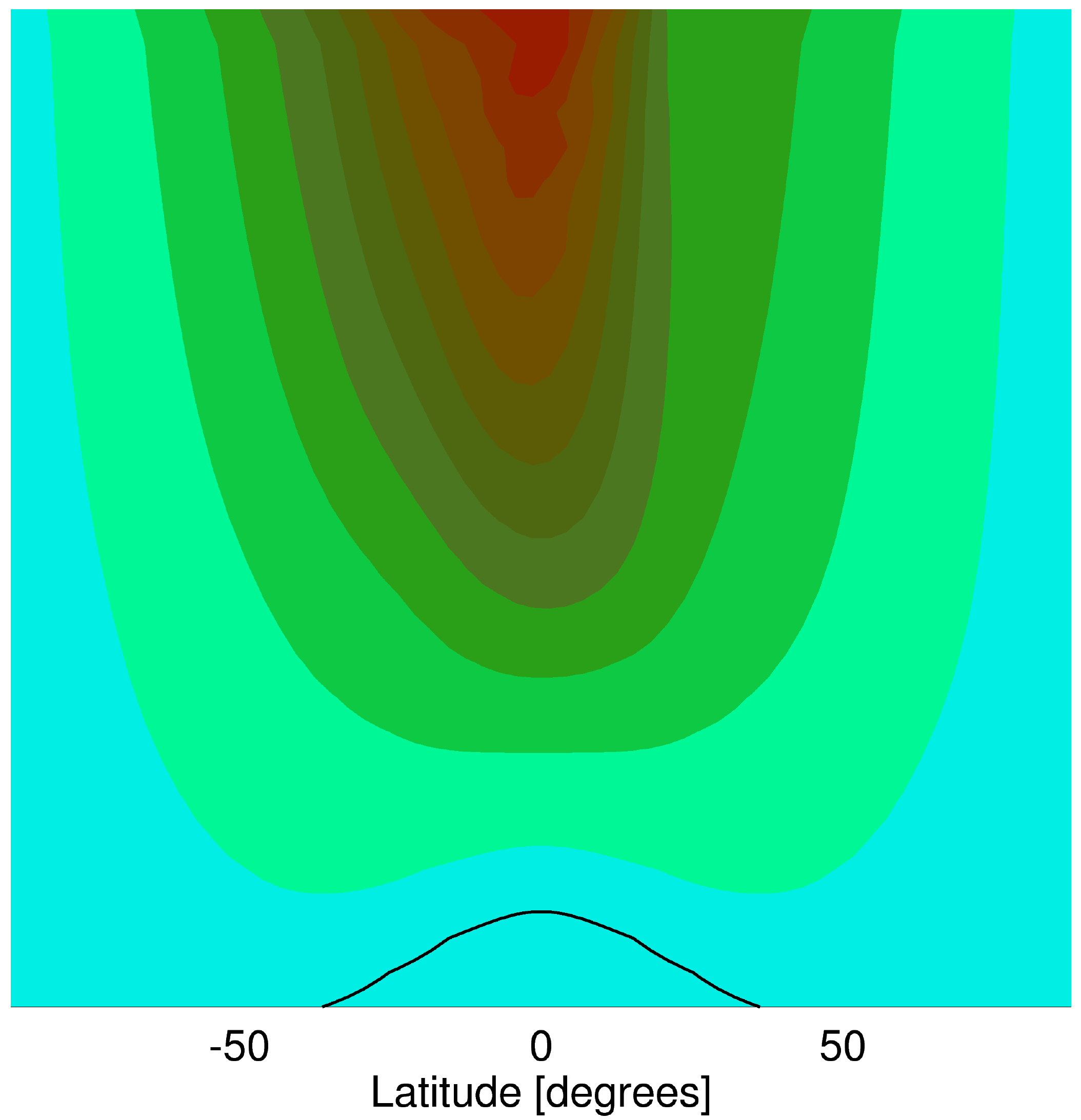}
\end{center}
\caption{The zonally averaged zonal (East-West) winds for models with obliquities of: 0\degr~(top left),
  10\degr~(top right), 30\degr~(third row from the bottom),
  60\degr~(second row from the bottom), and 90\degr~(bottom
  row).  The plots for $\psi=0\degr$ and $\psi=10\degr$ are 
  annual averages; the other plots are snapshots at:
  (Northern) vernal equinox, summer solstice, autumnal equinox, and
  winter solstice (left to right).  The colorscale is the same for
  each plot and the black line is boundary between Eastward
(positive) and Westward (negative) winds.}  \label{fig:uz}
\end{figure*}

Our models with relatively low obliquity (0\degr, 3\degr, and
10\degr) all exhibit atmospheric structures where the equator is
warmer than the poles and the circulation is eastward throughout most
of the atmosphere, dominated by a jet at high latitude in each
hemisphere.  The $\psi=0\degr$ and $\psi=3\degr$ models
are largely indistinguishable, so much so that we do not find it
worthwhile to present results from the 3\degr~model. Between the 0\degr~and
10\degr~models there are only small differences in wind speed and jet
width.  In Figure~\ref{fig:uz} we show plots of the zonally 
averaged zonal wind (in the East-West direction) for each model; the
plots for the 0\degr~and 10\degr~models are yearly averages,
since little-to-no seasonal variation is observed, while for the
higher obliquity models several snapshots are shown throughout the
planet's year.

For this hypothetical planet, we find that significant seasonal
differences emerge once the obliquity reaches 30\degr, with a
clear shift in wind patterns and temperature structure as the Northern
and Southern hemispheres receive more or less direct heating.  The
circulation pattern for the $\psi=30\degr$ model resembles the
patterns for lower obliquity models, with dominant eastward flow and
high latitude jets, but the jet structure is slightly distorted
between the two hemispheres, with a seasonal dependence.  The flow
patterns for the $\psi=60\degr$ and $\psi=90\degr$ models also 
 show a seasonal dependence, but the winds in these
 models are significantly different from the lower obliquity cases.
 For $\psi=60\degr$ there is a more even distribution between eastward and
 westward winds and the flow is characterized by an eastward jet that
 remains near the equator, but shifts up and down in latitude
 seasonally.  The 90\degr~obliquity model is dominated by westward
 flow, with a strong jet at the equator that shifts slightly in
 latitude throughout the year.  These planets are clearly in a different
 circulation regime than lower obliquity planets.  

\begin{figure}[ht!]
\begin{center}
\includegraphics[width=0.5\textwidth]{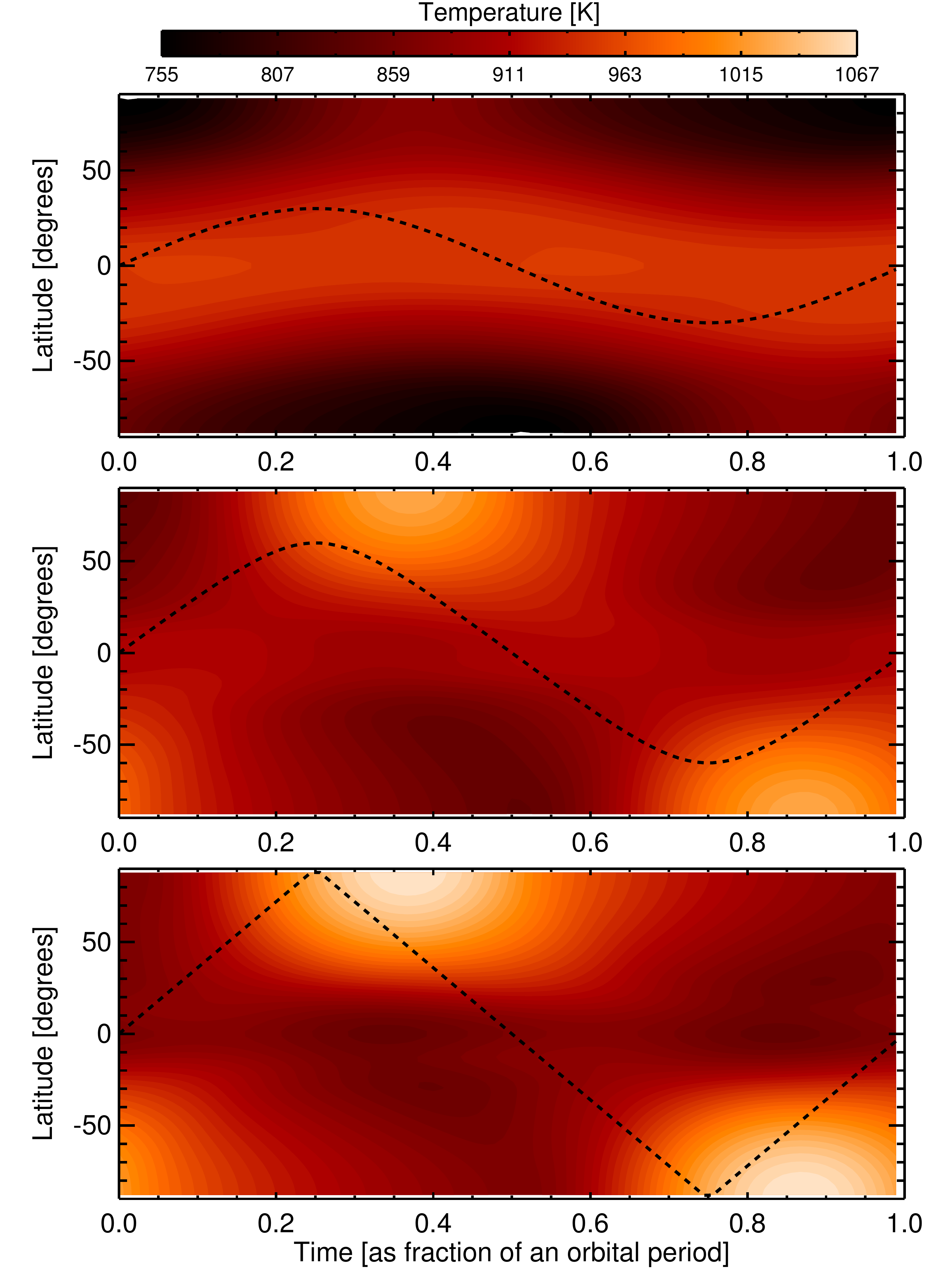}
\end{center}
\caption{The latitudinal temperature structure at the infrared
  photosphere (red contours, with color scale at top) and the latitude of the substellar
  point (black dashed line, Equation~\ref{eqn:pss}),  as a function of time throughout one
  orbit of the planet.  From top to bottom the panels show results for
  the models with 30\degr, 60\degr, and 90\degr~obliquity, respectively.  In all
  models the temperature structure lags the stellar irradiation
  pattern, with a timescale of $\sim$1/8 \Porb.} \label{fig:sstlat}
\end{figure}

In Figure~\ref{fig:sstlat} we show how the temperature
  structures of the $\psi \ge 30\degr$ models respond to the changing
  seasons.  We plot the temperature at the infrared photosphere for 
  each of these models as a function of time, throughout one orbit of
  the planet.  For each model we plot the latitudinal temperature
  structure by taking an azimuthal average (as is appropriate since
  our diurnal heating pattern results in no significant variation of
  properties with longitude).  On top of these temperature contours we
  plot a dashed black line to show the latitude of the substellar
  point as a function of time (Equation~\ref{eqn:pss}), so that we can
  compare the spatial and temporal response of the atmosphere to the
  changing irradiation pattern.

For the $\psi=30\degr$ model the temperature structure is
  similar to that of lower obliquity models, in that the equator
  remains warmer than the poles, but the latitude of maximum
temperature shifts above and below the equator with time.  The stark
change in atmospheric regime between the $\psi \le 30\degr$ and $\psi
\ge 60\degr$ models seen in the wind patterns (Figure~\ref{fig:uz}) is
also reflected in the temperature structures.  At most times during
the year one hemisphere is significantly warmer than the other,
such that the main temperature gradient is between the summer
  and winter hemispheres, with short times of relatively uniform
  temperature near the equinoxes.

From Figure~\ref{fig:sstlat} we can also see that the atmospheres
  of the $\psi=30\degr$, 60$\degr$, and 90$\degr$ models all respond
  to the changing irradiation pattern with a similar response time.
  This is observed as a lag in the latitude of the maximum
temperature, relative to the substellar latitude, by $\sim$1/8 \Porb.
This response time is about an order of magnitude longer 
than the radiative timescale at the infrared photosphere ($\sim$0.014
\Porb), evidencing a complex radiative and dynamical response of
the atmosphere as it transports energy from the
heating at the optical photosphere (where the radiative timescale is
also shorter than the response time, at $\sim$0.05 \Porb).

The focus of this paper is the observable consequences of non-zero
obliquities and so a more detailed analysis of these circulation
patterns and the shift between regimes will have to wait for future
work. The take-away point from this section is that seasonal effects
are insignificant for our models with obliquities $\leq10$\degr, but
influence the temperature structure at the infrared photosphere (i.e.,
what we would see in thermal emission) for models with obliquities
$\geq30$\degr. 

\section{The effect of obliquity on observable properties} \label{sec:obs}

The nature of any
observable seasonal variation will depend not only on the intrinsic
physical properties of the system (e.g., \Porb, $\psi$, stellar
flux, etc.), but also on the extrinsic viewing orientation of the observer
relative to the planet's axial tilt, which can be characterized by the
coordinates of the subobserver point.  Since we are considering
diurnally averaged models here, all subobserver longitudes are equal.
The subobserver latitude, \pobs, sets the viewing orientation and
remains constant with time; it is related to the obliquity and observed orbital phase
($\gamma$) as:
\begin{equation}
\sin \pobs = \sin \psi \sin (2 \pi [\gamma - t/\Porb]) \label{eqn:pobs}
\end{equation}
where $t=0$ is the same as in Equation~\ref{eqn:pss}. Transit is
defined to occur at an orbital phase of $\gamma=0,1$ and 
secondary eclipse is at $\gamma=0.5$ (assuming a transiting system and
a circular orbit).
Since the seasonal orientation of any system is
\emph{a priori} unknown, Equation~\ref{eqn:pobs} allows us to analyze
output from our models for fixed observer orientations (\pobs),
snapshots from the simulation ($t/\Porb$), or orbital phases.  A full phase
curve observation occurs over a planet's year ($\Delta t = \Porb$) and
requires a choice for the viewing orientation.  

For an eclipse mapping observation (at $\gamma=0.5$), we need to know not just the
subobserver latitude, but in which direction the pole of the planet is pointed.
Using the spherical law of cosines 
\citep[following the example of][]{Schwartz2016}, we find that the angle, $C$, through which the
Northern pole should be rotated (in a clockwise direction) around a line that passes from the center of the
planet to the observer (through the subobserver point), is given by:
\begin{equation}
\cos C = \cos \psi / \cos \pobs. \label{eqn:C}
\end{equation}
Cosine is an even function and the solution for $C$ could be a
positive or negative value; these correspond to the two points during
a planet's orbit when an observer might be oriented such that
secondary eclipse occurred at that value of \pobs.  From a
consideration of the orbital geometry (using Equation~\ref{eqn:pobs})
it is apparent that the positive solutions correspond to $-1/4 <
t/\Porb < 1/4$ and the negative solutions to $1/4 <
t/\Porb < 3/4$.

\subsection{Orbital thermal phase curves} \label{sec:pcurves}

In the simulated orbital phase curves presented here we assume that
our hypothetical planet is of interest because it transits and we choose
to view the system exactly edge-on.  This allows us to isolate the
observable consequences of non-zero obliquities, but for actual
systems these effects would be convolved with the effect of viewing at
an orbital inclination slightly less than exactly 90\degr. This will
be a minor effect and in practice the orbital inclination is
measurable from the shape of the transit curve and so the reference
frame could be adjusted to account for this angle.

\begin{figure*}[ht!]
\begin{center}
\includegraphics[width=0.75\textwidth]{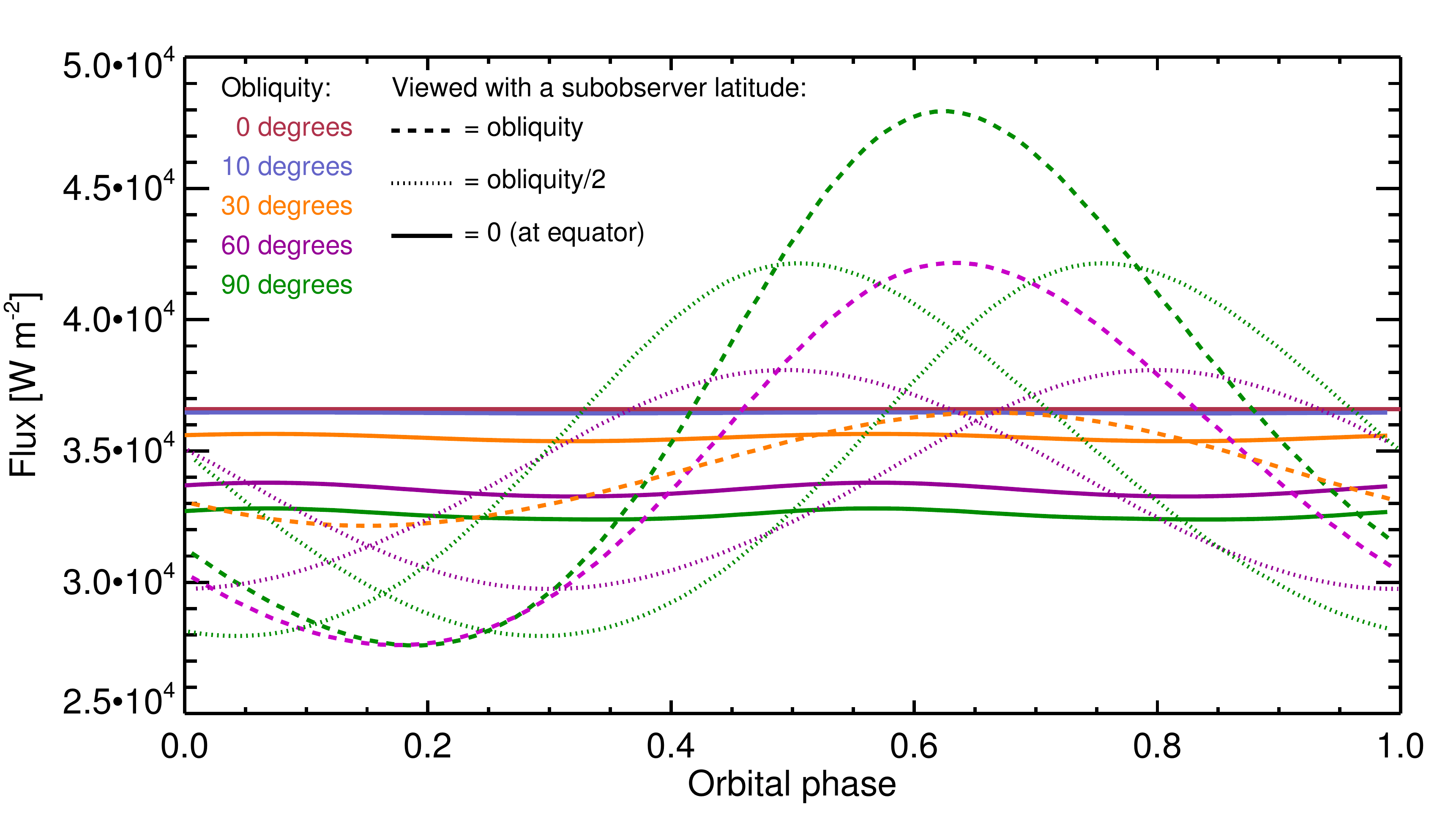}
\end{center}
\caption{Orbital phase curves for models with a range of obliquities.
  These curves show the amount of flux that the planet is emitting,
  from the hemisphere facing a distant observer, as a function of time
  throughout the planet's orbit.  Not only does the amount of flux
  differ between models with different obliquities (shown as different
  colors), but the observed flux variation from any particular model
  depends on the orientation with which the observer is viewing the
  planet (shown as different line styles).  Maximal variation would be
  seen if the axis of the planet is tilted toward the observer such
  that one hemisphere is primarily in view ($\pobs=\psi$), whereas the
  least variation is seen if the planet's axis is perpendicular to the
  observer such that equal parts of the Northern and Southern
  hemisphere are in view ($\pobs=0$).  In
  all cases the geometry of the system is assumed to be transiting;
  secondary eclipse would occur at the orbital phase 0.5 (when the
  observed planet flux drops to zero as it passes 
  behind its star, but not included in these curves).  The
  corresponding planet emission flux maps for each curve are shown in
  Figure~\ref{fig:fmaps} and full-orbit movies are available as
  supplemental files.} \label{fig:pcurves}
\end{figure*}

\begin{figure*}[ht!]
\begin{center}
\includegraphics[width=0.75\textwidth]{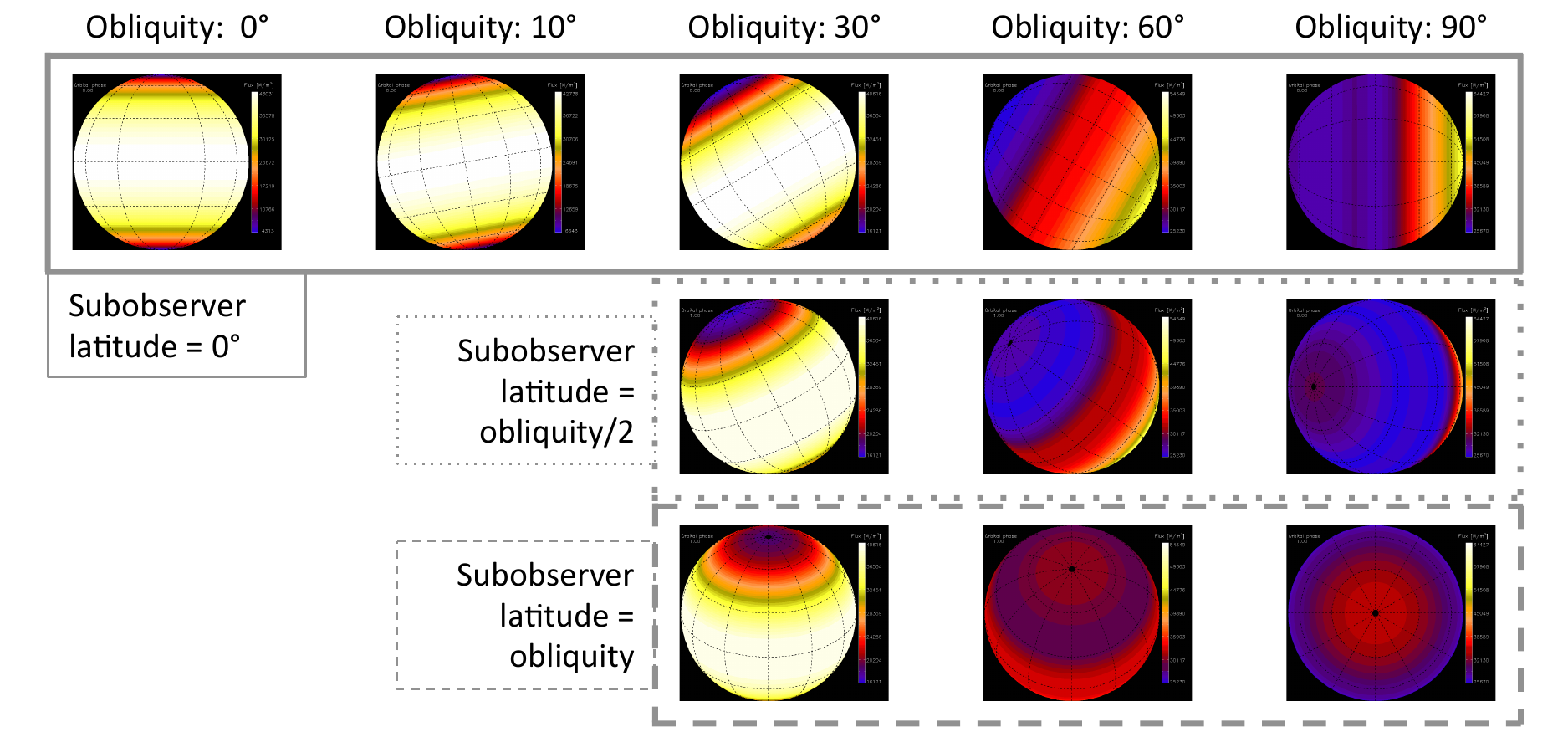}
\end{center}
\caption{Maps of the flux emitted from models with various obliquities
(from left to right, as labeled), and as observed with different viewing
orientations. In all cases the planet is assumed to be in a transiting
orbit and these images are for the time when $\gamma=0$, meaning that
the planet would be at the center of transit.  The top row show the
planet as it would look if the  
equator of the planet was aligned toward the observer. The middle row
show the planet as if the observer were seeing the planet at a
latitude equal to half of the planet's obliquity (so at 15$\degr$ for
the $\psi=30\degr$ model, 30$\degr$ for the $\psi=60\degr$ model, and
45$\degr$ for the $\psi=90\degr$ model). The bottom row shows the
planet oriented such that its rotation axis is pointed toward the
observer, so that the observer views a latitude equal to the
obliquity.  Movies showing the time variation of this flux pattern
throughout the planet's orbit, for all models and viewing orientations
represented here, are available as supplemental files.}\label{fig:fmaps}
\end{figure*}

Since we force the orbital inclination of our hypothetical system to
be exactly edge-on to the observer, this constrains the possible
viewing geometry such that the planet must be seen with a subobserver
latitude ranging from zero to $\pm \psi$ (and for the zero obliquity
model this means that the planet is also always viewed exactly along its
equator).  We calculate a set of simulated phase curves for each of our
  models by assuming various orientations (i.e. using different values
  for the subobserver latitude in Equation \ref{eqn:pobs}), and then
  integrate the flux emitted from the hemisphere facing the
  observer, as a function of time throughout one orbit, as the planet
  undergoes seasonal changes.  It is necessary to
  simulate a set of curves for each non-zero obliquity model because
  the random orientation of the observer relative to the system will
  directly affect the measured phase curve, as seen in
  Figure~\ref{fig:pcurves}.  Figure~\ref{fig:fmaps} shows maps of the 
  flux emitted from each of our models, oriented for the different
  viewing geometries used to produce the curves in
  Figure~\ref{fig:pcurves}.  Each map is for a snapshot of the planet
  at an orbital phase of zero; full-orbit movies showing the seasonal flux
  variation for the cases shown in Figure~\ref{fig:fmaps} are available
  as supplemental files.

If our hypothetical planet has zero obliquity, then its orbital phase
curve should be completely flat. Its temperature structure is
axisymmetric (due to its assumed Jupiter-like rotation rate
  and diurnally averaged stellar heating pattern) and it experiences
no seasonal variations, so the emitted flux is constant with time, as
seen in Figure~\ref{fig:pcurves}.   We can compare the flux level of
this curve to the flux level that we 
would expect if the heating from the star was evenly re-radiated from
the entire planet surface, ($F_{\mathrm{irr}}/4=3.4\times 10^4$ W
m$^{-2}$). We find that the flux from our $\psi=0\degr$ model is lower
than this global average, which is not surprising. Atmospheric
transport does not completely homogenize the 
equator-to-pole temperature difference from differential heating,
leaving the equator hotter than the global average, and we
preferentially observe this brighter region of the planet.\footnote{Note that the internal heat flux of
  the planet is several orders of magnitude less than the incident
  stellar irradiation and so largely irrelevant (see
  Table 1).}  If we could independently constrain the
obliquity of a planet to be zero, then the flux level of the orbital
phase curve would be a measure of the efficiency of equator-to-pole
heat transport, with values closer to the globally averaged flux
meaning more efficient transport.

In the case that it cannot be measured independently, the obliquity of
the planet has a degenerate effect with the equator-to-pole heat
transport on the observed flux level, for equatorial viewing geometries.
Figure~\ref{fig:pcurves} shows that---at equatorial viewing
orientations---a lower level of emitted flux is associated 
with a higher value of planetary obliquity.  Values less than the
globally averaged incident flux ($3.4\times 10^4$ W m$^{-2}$) are observed for planets with
obliquities $\ge60$\degr, a result of the
poles of these planets receiving more flux than the equator,
integrated over an orbit.  Figure~\ref{fig:sstlat} shows that
these high obliquity models have equatorial temperatures that
are always equal to or less than the planet's average
temperature. When the planet is observed along the equator we
preferentially see these cooler regions and so the average hemispheric
flux is decreased relative to the lower obliquity models.

The phase curves for equatorial viewing geometry are not completely
flat; there is some small amount of seasonal variation \citep[similar
to the semi-orbital variations in the blackbody planet models
of][]{Gaidos2004}.  Even though 
an observer would see an equal amount of each hemisphere,
Figure~\ref{fig:sstlat} shows that there are times of the year
when one hemisphere is much hotter than average and times when both
hemispheres are both only warm, leading to an integrated flux level that
shows a small amplitude seasonal variation.  However, this signal
would likely be lost in the noise of an actual phase curve measurement
(see Section~\ref{sec:measurement}) and so these curves are
effectively flat.

The take-away point from the solid curves in Figure~\ref{fig:pcurves} is
that if we observe a flat phase curve for the thermal emission from a
transiting planet, we are probably observing the planet along its
equator. If the measured flux level is above the globally
averaged value ($F_{\mathrm{irr}}/4=\sigma T_{\mathrm{star}}^4 [
R_{\mathrm{star}}/2a]^2$), then the planet's equator is hotter than
its poles and the obliquity is likely less than
54\degr.\footnote{Mathematically, $\psi=54$\degr~is 
the limit between illumination patterns where the equator or poles
receive more stellar flux, integrated over the planet's year.} In this
case the measured flux level is giving us
information about the efficiency of equator-to-pole heat transport.
If the observed flux is below the globally averaged value, then the
obliquity is likely greater than 54\degr~and our results show that
lower levels correspond to higher obliquities.  An important caveat to
this is that the albedo of the planet may be unknown, in which case
the globally averaged flux value would need to be adjusted down by an
unknown amount, confusing the interpretation of a flat phase curve.

The set of flat (or effectively flat) solid phase curves shown in
Figure~\ref{fig:pcurves} are for the particular viewing 
orientation of the subobserver point being along the planet's
equator.  Plotted as dashed lines in Figure~\ref{fig:pcurves} are the other extreme
viewing orientation, where the planet is observed at the maximum
possible subobserver latitude (for a transiting system), $|\pobs|=\psi$.  For these cases we
observe the maximum variation in emitted flux, since the full seasonal
response of a hemisphere is preferentially in view.  Without spatially
resolving the planet from the star we cannot know the orientation of
the orbital angular momentum vector, leading to a North-South
degeneracy in our observed properties.  We do not know whether our
subobserver latitude is positive or negative (Northern or Southern),
but since both hemispheres respond to the seasonal cycle
symmetrically, it is only the absolute value of \pobs~that is relevant anyway.

For the $\psi=10$\degr~model, which does not show significant
seasonal variation, the $\pobs=\pm 10$\degr~viewing 
orientation results in a slightly diminshed flux level. This is simply
explained by the higher latitudes on the planet being cooler than the
equator and so shifting those latitudes to greater visibility
diminishes the disk-integrated flux observed.  For the models that do
exhibit seasonality ($\psi=30,60,90$\degr) the amplitude of
variation increases for increasing obliquity, while the orbital phase of
the maximum flux is the same for all models.  This trend of increasing
variation with obliquity was also found for Earth-like planets in
\citet{Gaidos2004}, and their results can also be used to make 
sense of the constant phase we see for maximum flux, since the thermal
inertia of the atmosphere is the same for all models.

When we observe the planet at
$\pobs=\psi$, then according to Equations~\ref{eqn:pss}
and~\ref{eqn:pobs}, the solstice for the hemisphere facing us occurs
at $\gamma=0.5$ and so the phase difference between this and the peak
flux is a measure of the atmospheric response time to the changing
irradiation pattern.  This agrees with the peak of the curves in
Figure~\ref{fig:pcurves} being at 0.5 $+\sim$1/8 $\approx 0.625$,
matching the timescale for atmospheric response determined above (see
Section~\ref{sec:circ}).  

In the bottom plot of Figure~\ref{fig:pcurves} we show intermediate
viewing angles for the highest obliquity models
($\psi=60,90$\degr), those with the strongest seasonal variation.
When $\pobs \neq \psi$ two solutions exist for the orientation of the
observer relative to the system.  When $\pobs=0$\degr, as in the
solid curves of Figure~\ref{fig:pcurves}, these two solutions produce
overlapping phase curves because of hemispheric symmetry, but at
intermediate values they do not, as is evident in the $\pobs=\psi/2$
curves shown as dotted lines.  For one solution the planet is
observed such that solstice in the primarily observed hemisphere
occurs before $\gamma=0.5$ and in the other solution it occurs after
$\gamma=0.5$, resulting in phase shifts in either direction.  The
phase at which peak flux occurs depends on a combination of the
atmospheric response timescale and the subobserver latitude, with a
positive or negative shift depending on which orientation we are
observing from, relative to the seasonal cycle.

In general we have learned that the orbital phase curves of emitted
light from this hypothetical planet contain information about the
planet's intrinsic properties of obliquity, and either equator-to-pole heat
transport (for low obliquity), or the atmospheric timescale for
seasonal response (for high obliquity).  The expression of these properties is regulated by
the viewing orientation of the system, which could make it difficult
to disentangle the intrinsic information without some other
independent constraint on the viewing or inherent properties
\citep[confirming the conclusions of][]{Gaidos2004}.
However, if the planet obliquity or viewing geometry were known, then
we could constrain atmospheric properties through the type of analysis
demonstrated here. In the next Section we discuss another type of
measurement that could be done instead, or could be used to break the
degeneracies in the phase curves.

\subsection{Eclipse maps} \label{sec:emaps}

Orbital phase curves have been the primary means by which we have
measured spatial information about exoplanets but, as we just saw,
this type of observation should become less informative (and much more
time-consuming) for planets on longer orbital periods. Conveniently,
another type of 
observational technique, known as eclipse mapping, should be
increasingly useful for longer period planets.  In this method, the
detailed shape of the flux curve as the planet passes into secondary
eclipse (ingress) or comes out from behind the star (egress), can be
inverted to construct a two-dimensional map of 
the planet's dayside brightness pattern
\citep{Williams2006,Rauscher2007b,deWit2012,Majeau2012}.  The
advantages of using this method for longer period planets are:
\begin{itemize}
\item It only requires observation during the time surrounding
  secondary eclipse, requiring much less telescope time than a full
  orbit (which would be 10 days for our hypothetical planet).
\item The slower orbital speed of the planet means that the ingress
  and egress take longer, providing a better opportunity to
  characterize the detailed shape of this part of the light curve.
\item The two-dimensional information retrieved by this method can
  constrain multiple properties of the planet.  In addition to the
  magnitude of the equator-to-pole or North-to-South temperature
  gradient, the orientation of the flux gradient can reveal
  information about the planet's obliquity.
\end{itemize}

The primary measurement made during secondary eclipse is the total
decrement in light when the planet is blocked from view, behind the
star. The total depth of the eclipse then gives us the integrated flux
from the dayside hemisphere of the planet. From
Figure~\ref{fig:pcurves} we can see that the amount of planetary flux
that would be eclipsed by the star (the value of each curve at an orbital phase of 0.5) can
depend strongly on the obliquity of the planet and the viewing
orientation, but degeneracies exist such that these two properties
cannot be uniquely retrieved.  For example, in 
Figure~\ref{fig:pcurves} some of the curves for the $\psi=60\degr$
model would be observationally indistinguishable from 
some curves for the $\psi=90\degr$ model at 0.5
orbital phase; many more overlapping possibilities exist for the
full range of obliquities and viewing geometries.  The secondary
eclipse depth alone does not provide a strong constraint on the planet's
obliquity. 

\begin{figure*}[ht!]
\begin{center}
\includegraphics[width=0.235\textwidth]{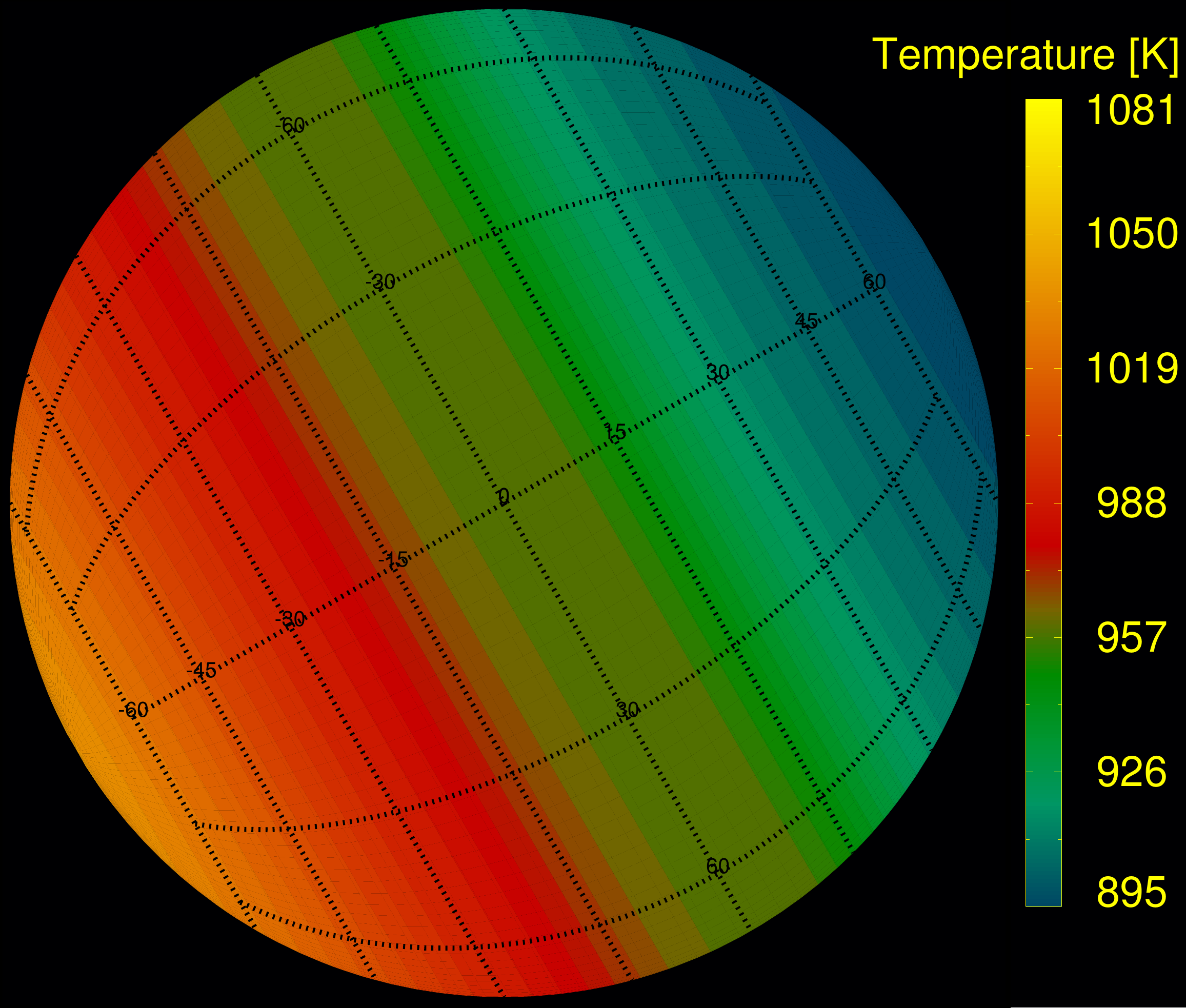}
\includegraphics[width=0.235\textwidth]{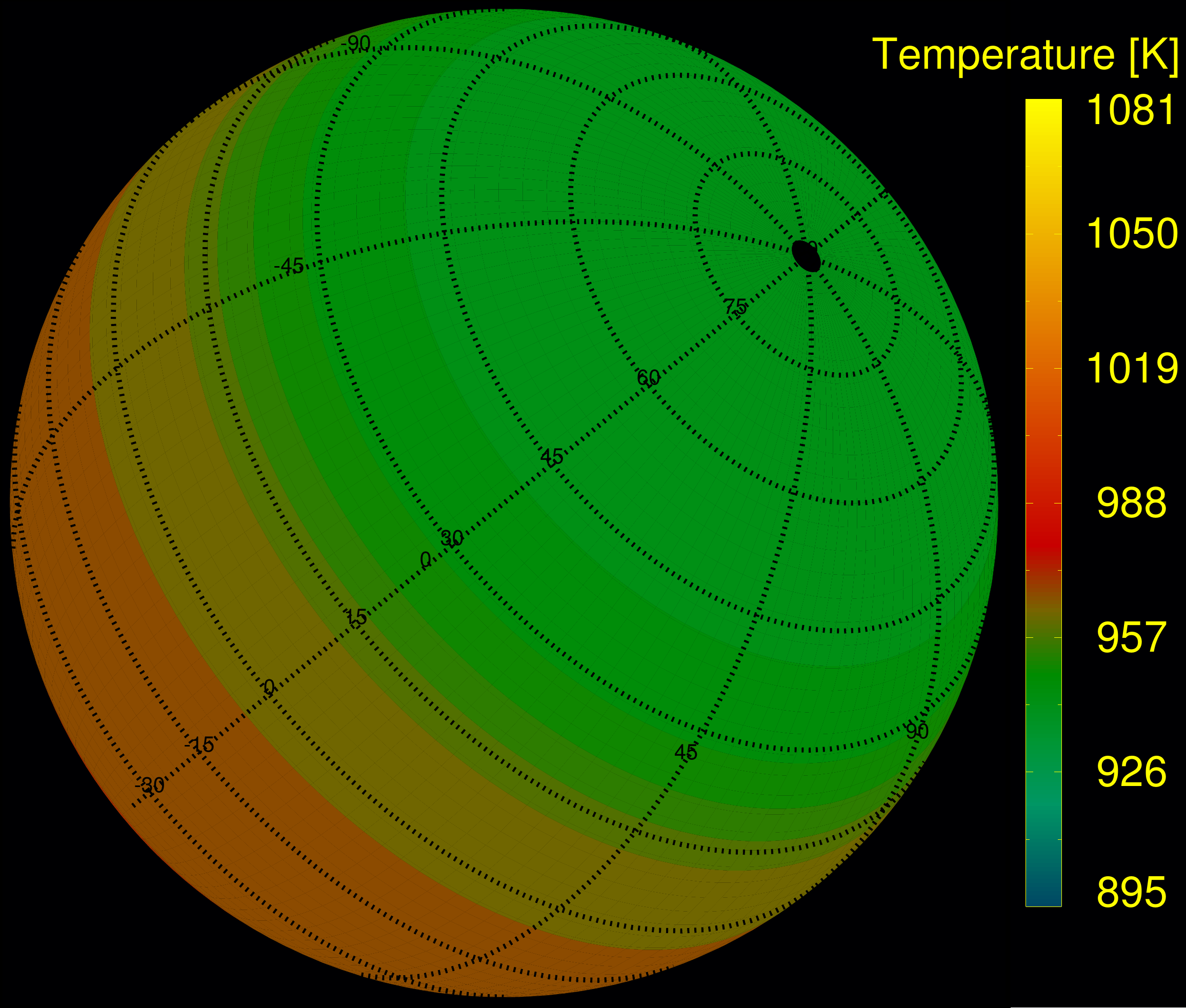}
\includegraphics[width=0.235\textwidth]{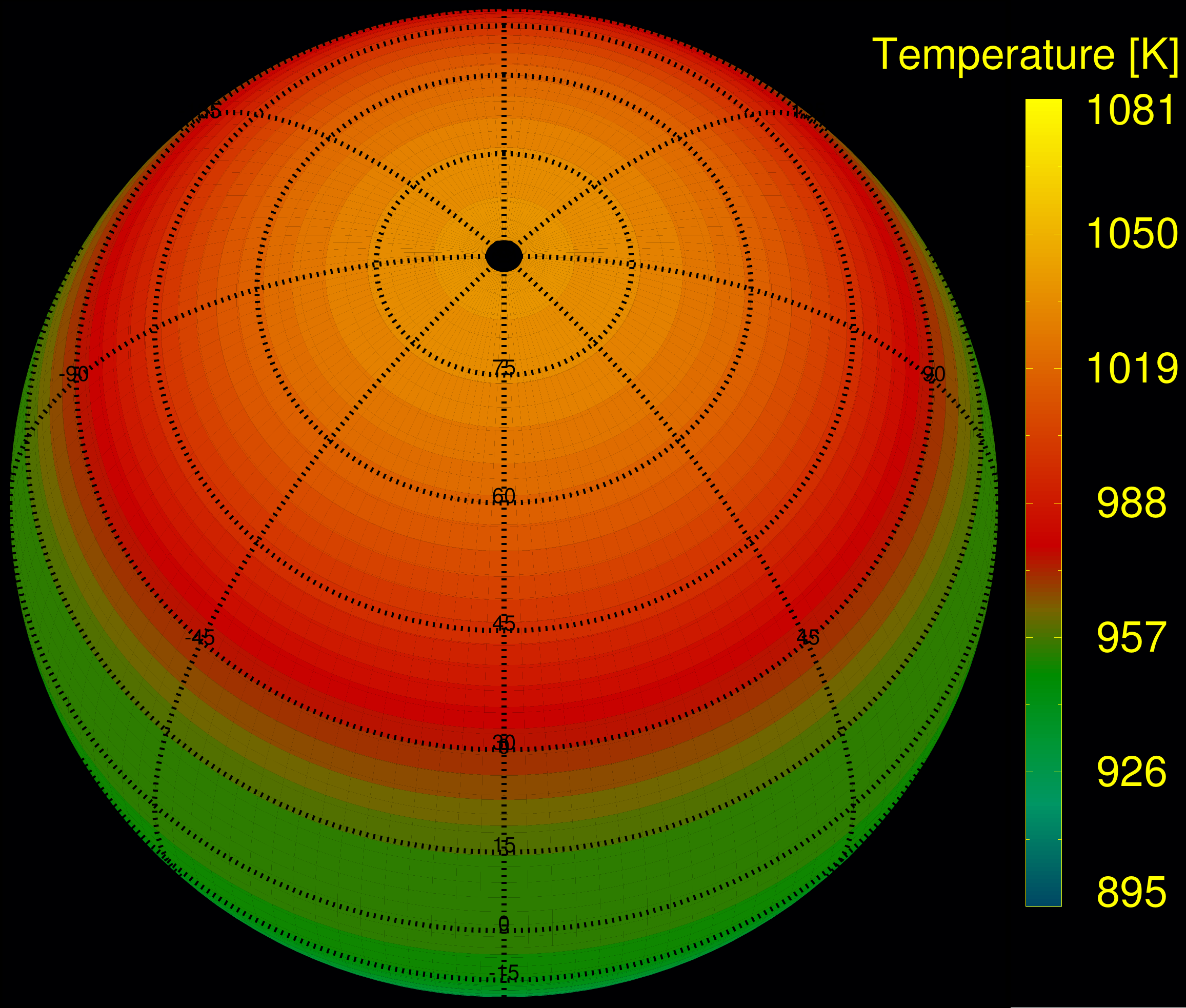}
\includegraphics[width=0.235\textwidth]{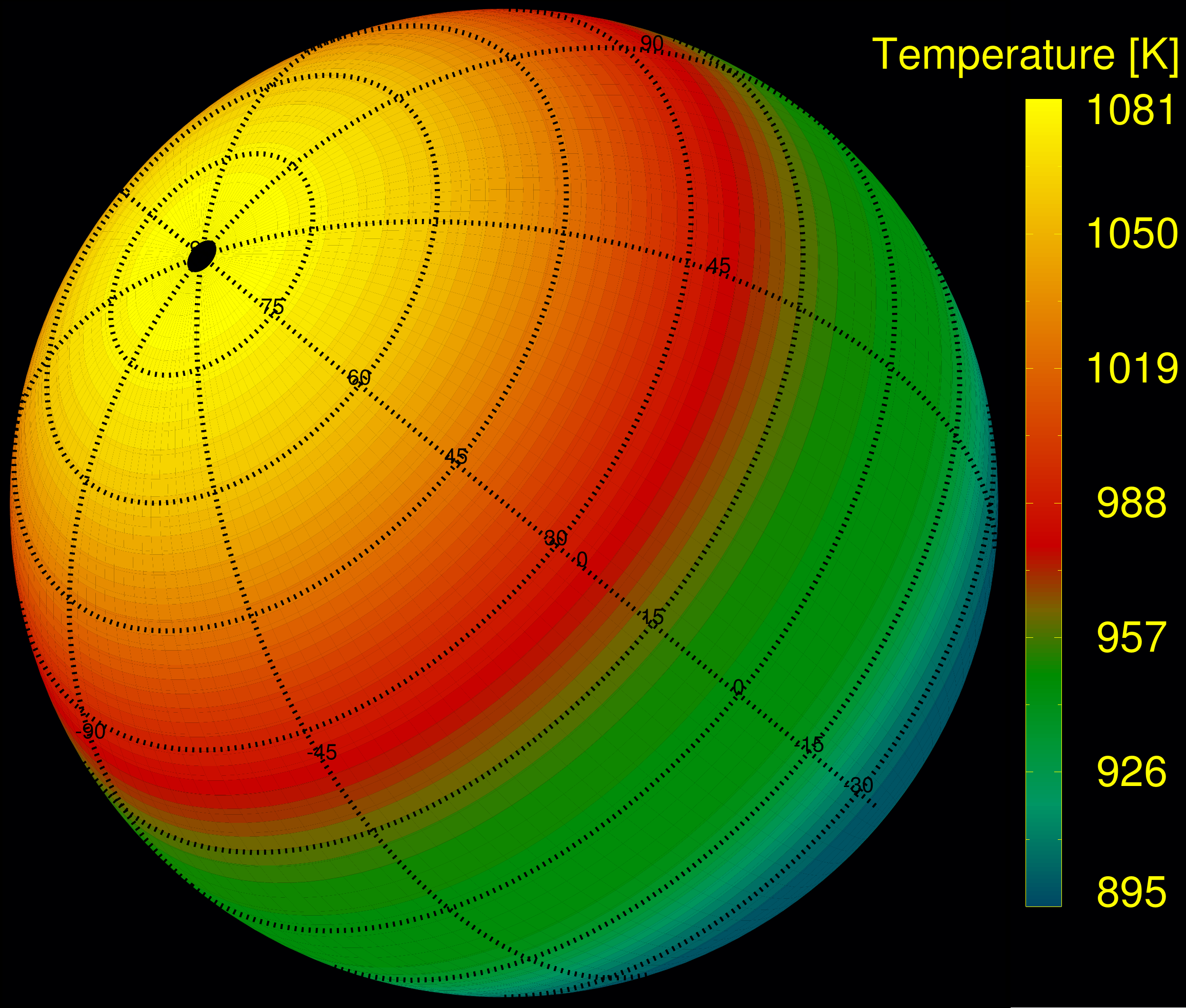} 
\end{center}
\caption{Temperature structure at the infrared photosphere, for the
  $\psi=60$\degr~model, at $t/\Porb=0$, $1/8$, $2/8$, and $3/8$
  (from left to right). In each plot the axis is oriented to the perspective of
  an observer who would see the planet go into secondary eclipse at
  that time, assuming that the angular momentum vector of the orbit is
  pointed up on the page.} \label{fig:t60}
\end{figure*}

Eclipse \emph{mapping} provides the additional information required to
discriminate between the temperature structures of low- or
high-obliquity planets, in the case where they have the same hemispherically
integrated flux.  In Figure~\ref{fig:t60} we show the
temperature map at the infrared photosphere for the 60\degr~obliquity
model, at several snapshots throughout its orbit.  For each snapshot
we have oriented the planet image to the perspective of an observer
who happens to see the planet in secondary eclipse at that point in
its orbit (using Equations~\ref{eqn:pobs} and \ref{eqn:C}; the viewing
orientation changes between each snapshot). From this
careful accounting for the geometry of secondary eclipse at various
viewing orientations, we can simulate different possible eclipse
mapping measurements. In Figure~\ref{fig:t60} we only show snapshots
from the first half of the orbit because the 
images from the remainder of the planet's year are mirror-images,
flipped vertically: the $t/\Porb=4/8$ snapshot is the flipped image of
the $t/\Porb=0$ snapshot, the $t/\Porb=5/8$ snapshot the flipped image
of $t/\Porb=1/8$, and so on.

\begin{figure}[ht!]
\begin{center}
\includegraphics[width=0.5\textwidth]{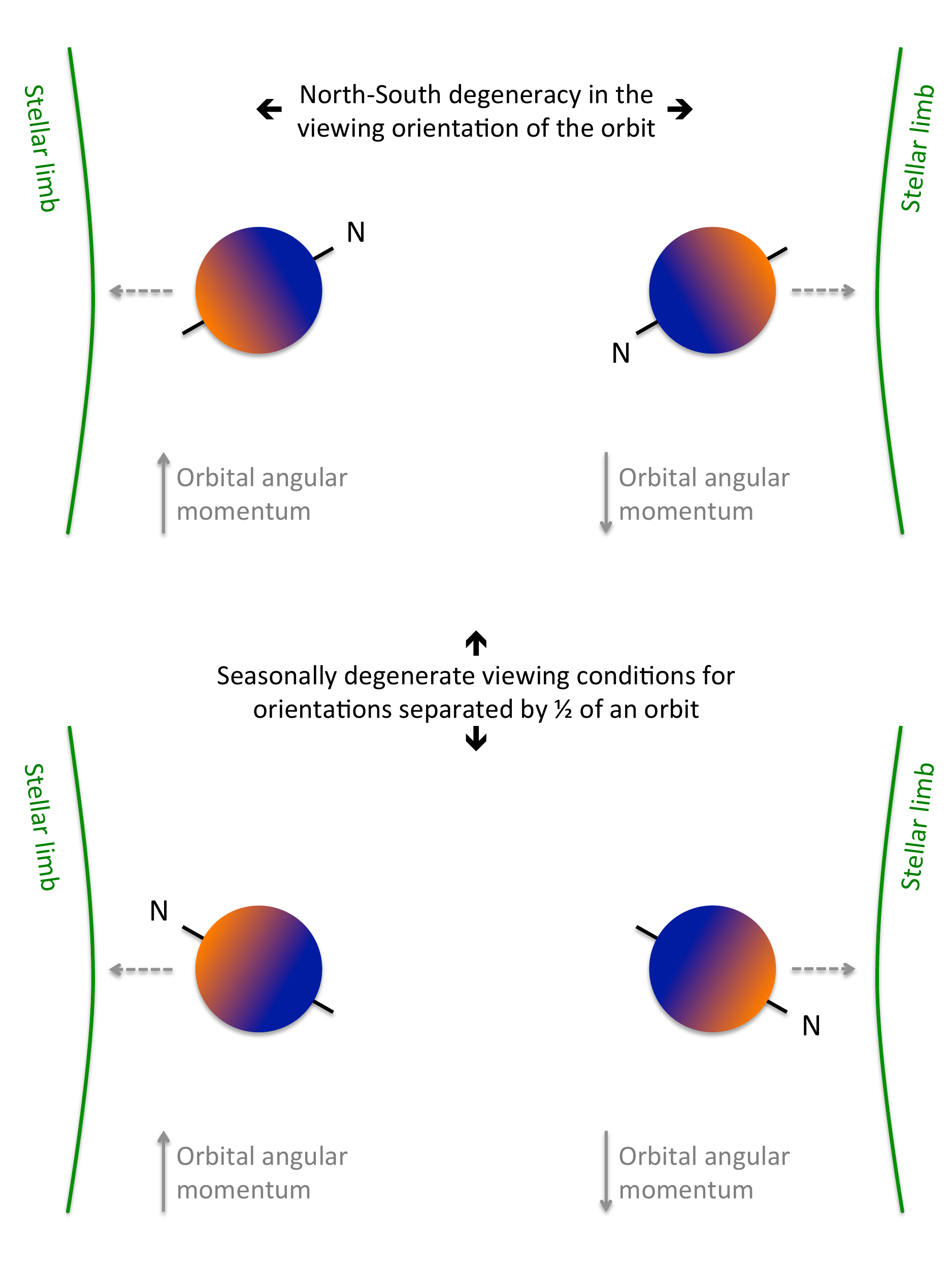}
\end{center}
\caption{A diagram of the degeneracies that result from the
unknown viewing orientation of the system.  As distant observers, we cannot
tell which hemisphere is the planet's North or South, nor can we
tell Northern summer from Southern summer.} \label{fig:eclipsediag}
\end{figure}

There are two important degeneracies that influence the
orientation of the brightness structure that can be measured by
eclipse mapping: the unknown North-South orbital orientation of the
system, and a seasonal hemispheric degeneracy.
Figure~\ref{fig:eclipsediag} shows schematic examples of these
degeneracies.  Since we are not spatially resolving the planet from
the star, but rather using the time domain to separate out its signal,
we do not know the orientations of the orbital or rotational
angular momentum vectors relative to us, the latter of which defines
which hemisphere is the Northern one.  We may be able to determine that one hemisphere is hotter than
the other, but cannot label it as ``North'' or ``South''.  This
degeneracy was also discussed above in Section~\ref{sec:pcurves}, as
it relates to the phase curves only being sensitive to the absolute
value of the subobserver latitude.

The second degeneracy diagrammed in Figure~\ref{fig:eclipsediag} is a
physical degeneracy in the seasonal response between the two
hemispheres.  There are no sources of North-South asymmetry for our
idealized hypothetical planet and so we see
identically changing heating and cooling patterns in each hemisphere's
temperature structure.  This explains why times separated by one half of
an orbit have mirror-imaged temperature patterns.  Since the
North-South viewing degeneracy also exists, this means that we cannot
discriminate between these seasonally identical conditions.  For the
example diagrammed in Figure~\ref{fig:eclipsediag}, we cannot differentiate
between a planet that is passing into eclipse in Northern winter or
Northern summer, because in either case the hotter hemisphere will be
eclipsed first.  This degeneracy is also implicitly found in
Figure~\ref{fig:pcurves}, in that each curve is actually two
overlapping predictions: the $+\pobs$ and $-\pobs$ solutions with
times separated by half an orbit.

\begin{figure*}[ht!]
\begin{center}
\includegraphics[width=0.65\textwidth]{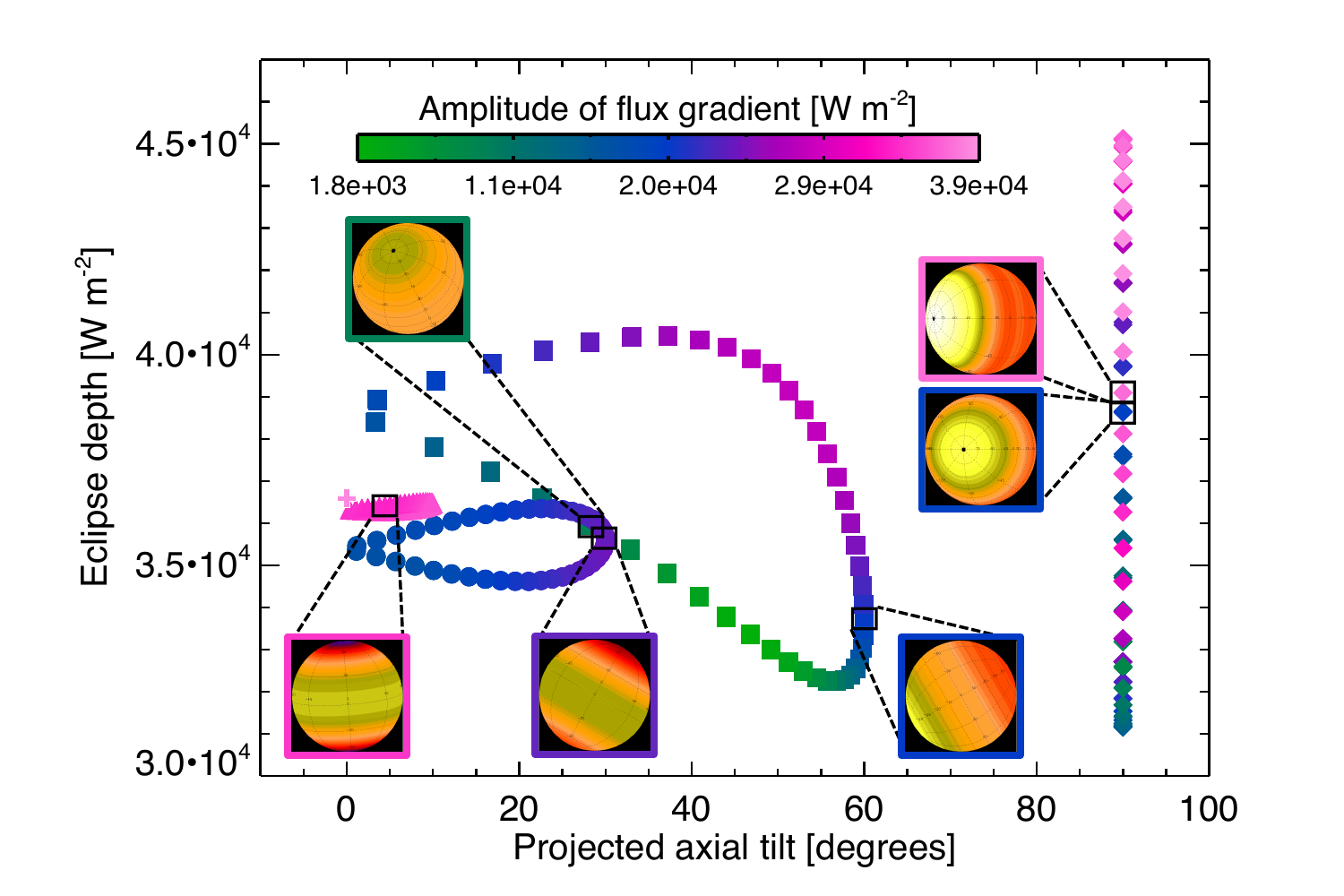}
\end{center}
\caption{The properties of a planet measurable by eclipse mapping: the
total flux from the observed hemisphere of the planet (eclipse depth),
the gradient in flux across the planet disk 
(amplitude in color), and the orientation of the planet's axis
relative to the stellar limb, assuming the rotation is aligned with
the flux gradient (projected axial tilt). Each point corresponds to
the properties of a
map that could be measured for models with obliquity equal to
0\degr~(cross), 10\degr~(triangles), 30\degr~(circles),
60\degr~(squares), and 90\degr~(diamonds), depending on the
orientation of the observer relative to the planet's orbit.
Also shown are the flux maps corresponding to several
  representative datapoints, oriented as they would be observed in
  secondary eclipse, with the stellar limb moving horizontally across
  the page.  The color gradient for these maps is the same as in
  Figure~\ref{fig:fmaps} (white/yellow is the brightest, red is
  dimmer) and all maps share the same scale so that they can be
  directly compared to each other.} \label{fig:einfo} 
\end{figure*}

Even with those degeneracies, there still are several important pieces
of information that we can measure with eclipse mapping. We calculate
these from each of our models and plot them in Figure~\ref{fig:einfo}:
\begin{enumerate}
\item  On the vertical axis we plot the secondary eclipse depth, or disk-integrated
  flux emitted from the hemisphere facing the observer.
\item The color of each point shows the amplitude of the flux gradient, meaning the
  difference in flux between the brightest and dimmest regions of the
  planet, for the observed hemisphere.
\item On the horizontal axis we plot the projected axial tilt, meaning the
  orientation of the flux gradient relative to the direction of the
  motion of the stellar limb. Specifically, we calculate the projected
  angle of the observed flux gradient and take the difference between
  this an the planet's orbital angular momentum (which is
  perpendicular to the direction of motion of the stellar limb). Due
  to the degeneracies discussed above and diagrammed in
  Figure~\ref{fig:eclipsediag}, we only calculate the absolute value
  of this tilt (i.e., not whether it is tipped clockwise or
  counterclockwise) and only characterize it with values $\leq90\degr$
  (i.e., do not differentiate between the Northern or Southern
  hemisphere).
\end{enumerate}
The third property listed, the projected tilt of the flux gradient, is the
mostly directly observable indication of the planet's obliquity. This
is because our axisymmetric models always have flux gradients aligned
with the rotation axis and so we are effectively measuring the
projected obliquity. This will, however, be a function of the
(arbitrary) viewing orientation of the seasons relative to our
line-of-sight and so may take on a range of values. Results for each
of our obliquity models (plotted using different symbols) were
calculated at evenly spaced snapshots in time throughout 
the planet's orbit, in order to explore the possible range of observed
properties.  In Figure~\ref{fig:einfo} we also choose several
representative datapoints for the different obliquity models and show
the planet flux map that corresponds to each of those cases.

The zero obliquity model has no seasonal variation and its temperature
structure remains unchanging throughout the year.  Its eclipse depth,
flux gradient, and projected axial tilt will always be the same value,
independent of viewing orientation.  These values are all nearly
degenerate with the results for the $\psi=10\degr$ model, whose
temperature structure is very similar and shows only minimal seasonal
variation.  The only slight difference is that the project tilt for
this model could be observed as anything between 0\degr~and 10\degr,
depending on the orientation of the observer relative to the seasonal
cycle.  It is unlikely that observations could discriminate between
these low obliquity cases for our hypothetical planet.

When we compare the predicted eclipse mapping data for models with
higher obliquities and stronger seasonal variation, we do see
observable differences.  There are some times during its orbit when
the $\psi=90\degr$ model would be observed to have the same
eclipse depth and flux gradient amplitude as the $\psi=0\degr$ model, but
its projected axial tilt is always at 90\degr~(always aligned with
the direction of motion of the stellar limb and so perpendicular to
the orbital angular momentum); this would be a clear way to
discriminate between these two extremes in obliquity.

Another piece of information from eclipse mapping that would help to differentiate between planets with
obliquities of 0\degr~or 90\degr~is that the flux gradient for the
0\degr~case is peaked at the equator and drops off toward the
poles, while the 90\degr~case has a North-South gradient that
stretches across the whole planet.  This type of spatial information
is retrieved in eclipse mapping observations, although we have not
quantified it in our analysis here.  Nevertheless, it is worth
emphasizing that the planets with equators hotter than their poles
(our $\psi=$0\degr, 10\degr, and 30\degr~models) 
should be observed to have a different pattern of emission than
planets where one hemisphere is hotter and brighter than the other
(our $\psi=$60\degr~or 90\degr~models).  However, this will
depend some on the viewing orientation; for example, a high-latitude
viewing orientation for a planet with a hotter equator and cooler
poles could perhaps mimic the hemispheric flux difference of a higher
obliquity planet, depending on the level of detail that can be
retrieved with eclipse mapping.

Our models with obliquities of 30\degr~and 60\degr~show
intermediate eclipse-mapping properties in Figure~\ref{fig:einfo}.
They both have possible projected axial tilt values from zero up to
their full obliquities, but observing a non-zero tilt only constrains
the possible obliquities to be greater than or equal to that
value.\footnote{Only for the $\psi=0\degr$ and 90\degr~cases is the
  projected tilt always exactly equal to the obliquity.} However, here
we can see eclipse mapping's power to constrain a planet's
obliquity. While models with different obliquities could 
produce the same the same projected axial tilt, the same eclipse depth,
and/or the same flux gradient amplitude, these properties can
generally be used in combination to break degeneracies between the
different possible planet obliquities.  In the next Section we estimate the ability of current
and future instrumentation to perform precise enough measurements to
be able to constrain a warm Jupiter's obliquity.

\subsection{Current and future instrument sensitivity} \label{sec:measurement}

We have predicted two types of observations that could potentially be
used to discern a planet's obliquity: orbital phase curves and eclipse
maps. Here we compare the signals predicted by our models to the
instrumental sensitivity of current and future missions.  We use rough
estimates, since our modeled planet is hypothetical and we have to
wait for a specific system to be targeted before more detailed
estimates can be made.  First we go through the exercise of estimating
  signal-to-noise assuming that the planet orbits a bright star and
  that we can achieve photon-limited precision, before a discussion of
  how to scale to less optimistic circumstances.

\subsubsection{Predicted planet-to-star flux ratio}

We already have predictions for the flux emitted from the planet, as a
function of obliquity and viewing orientation (Figures~\ref{fig:pcurves} and \ref{fig:einfo}),
but the actual observable is the planet-to-star flux ratio, which we
estimate as follows.  The double-gray radiative transfer scheme used in our
circulation model separates
the optical absorbed stellar light from the thermally re-emitted
infrared light, with the result that the predicted emission shown
above is a general ``infrared'' emission and in fact encompasses all emission
from the planet \citep[for details, see][]{Rauscher2012b}.  If we view
this in a bolometric sense, then the 
planet-to-star flux ratio is $\sim6\times10^{-4}$, based on an average
planetary emission of $\sim3.6\times10^4$ \Wms~(from
Figures~\ref{fig:pcurves} and \ref{fig:einfo}) and assuming Sun-like
properties of the host star.

In comparing our predictions to specific instruments, it is important
to consider the strong wavelength dependence of the planet-to-star flux
ratio.  Assuming Planck functions for the planetary and stellar
emission, their flux ratio at some wavelength can be calculated as:
$(R_p/R_s)^2\times B(\lambda,T_p)/B(\lambda,T_s)$.  Using the planet's
equilibrium temperature of $\sim$880 K, which is close to the average
temperature at the planet's infrared photosphere
(Figure~\ref{fig:sstlat}), we can estimate the planet-to-star flux
ratio for Hubble (WFC3 G141, 1.4 micron) as $\sim 4\times10^{-7}$ and for Spitzer
(IRAC1, 3.6 micron) or JWST (NIRCam F356W, 3.5 micron) as $\sim 1\times10^{-4}$.

\subsubsection{Estimates of precision}

We base our precision estimates on the values from Table 3 of
\citet{Cowan2015}, which assume the photon-counting limit for a target
20 pc away around a 5000~K host star.
We adjust these predictions from an assumed integration time of 1 hour
to the $\sim$4 hour length of the secondary eclipse for our hypothetical
planet ($=2R_s/[2\pi a/\Porb]$).  Since our various measurements are
all differential (in- vs. out-of-eclipse, amplitude at one phase
vs. another) and so require the comparison of two different flux-level
measurements, we apply a $\sqrt{2}$ penalty and obtain optimistic
precisions of 1.6 ppm, 11 ppm, and 1.6 ppm (parts per
million) for Hubble (WFC3 G141), Spitzer (IRAC1), and JWST (NIRCam
F356W), respectively.   We can only make educated guesses as
  to JWST's actual performance until it begins operations, but based
  on lab tests of its detectors (noise floor estimates, quantum
  efficiencies, etc.), as well as previous
  experience with Hubble and Spitzer, we expect that JWST will achieve
  excellent photometric precision for exoplanet transits and eclipses
  \citep{Beichman2014}.

\subsubsection{Detection of secondary eclipse} 

The largest, easiest signal to detect is that of the secondary eclipse
of the planet. Using the planet-to-star flux ratios and precisions estimated in the
previous two sections, we expect that with a \emph{single} observation
  (assuming photon-limited precision) the secondary eclipse of our
hypothetical warm Jupiter would not be detected
by Hubble, but could have been detected by Spitzer at 9-$\sigma$,
and should be detectable by JWST at 80-$\sigma$.

This optimistic prediction implies that JWST should be able to measure
the eclipse depths plotted in Figure~\ref{fig:einfo} to a precision of
$\sim0.2\times10^4$ \Wms~at 3-$\sigma$.  The eclipse depths of the
models with $\psi=0\degr$ and 10\degr~will be
indistinguishable, the lowest possible values of the $\psi=30$\degr~
model may be barely distinguishable from the lower obliquity cases,
and much of the range of values for the $\psi=60$\degr~or
90\degr~models should be easily distinguishable.  Since these are
single-eclipse estimates, we could further improve the
precision by collecting multiple observations, since we see no
significant orbit-to-orbit variability on the planet.

\subsubsection{Detection of variation in orbital phase}

According to Figure~\ref{fig:pcurves}, the amplitude of orbital flux
variation we should observe depends both on the planet's obliquity and
our viewing orientation, with the (\emph{a  priori} unknown) viewing
orientation having a strong influence on the phases of maximum and
minimum flux. Since the orbital period is 10 days, it would be
observationally very expensive to monitor a 
continuous, full orbit of the system and so a few shorter integration
observations would probably be preferred \citep{Krick2016}.  If there was already a
secondary eclipse observation, that would help to constrain possible
combinations of obliquity and viewing orientation, and help to
intelligently schedule snapshots throughout the orbit.

We will side-step the issue of timing and instead just assume that
there is one 4-hour observation near the peak in flux and one 4-hour
observation near the minimum, in order to determine what level of
variation could be measured.\footnote{Four hours is $\sim$0.02 in
  phase and so we would not expect much seasonal variation in thermal
  emission during that integration time.}  From our estimates of
signal and precision, we 
determined above that JWST should be able to measure flux differences
of $\sim0.2\times10^4$ \Wms~for this planet at 3-$\sigma$.  This means
that all of the phase curves for equatorial viewing geometries (the
solid lines in Figure~\ref{fig:pcurves}) would be considered flat, as
would the maximally variable curve for the $\psi=10$\degr~model.
All of the rest of the curves shown in Figure~\ref{fig:pcurves}
(non-equatorially viewed models with $\psi \ge 30$\degr) would have
measurable variation, at the 9- to 40-$\sigma$ level with JWST.

\subsubsection{Detection of flux gradient via eclipse mapping}

The detection of a gradient in flux across the planet's disk can be
measured during the ingress and egress times of secondary eclipse,
given a high enough precision.  This limits the integration time of
the observation to the time that it takes the planet disk to move
behind the star (or re-emerge), $2R_p/(2\pi a/\Porb)=0.4$ hours.
Whereas for the eclipse depth measurement we assume an integration
time of 4 hours and apply a $\sqrt{2}$ penalty for comparison with
a baseline measured out of eclipse, here all of the information is
contained within the times of ingress and egress.  We thus use only
0.4 hours as the integration time, instead of 0.8 for the full ingress
plus egress, and still apply the $\sqrt{2}$ penalty for a differential
measurement.  We could simplistically think of this as a comparison in
flux between the first/second half of ingress/egress and the
second/first half of ingress/egress. While the details of eclipse
mapping are actually more complex
\citep{Williams2006,Rauscher2007b,deWit2012,Majeau2012}, this method
still relies on comparing the differential flux during some time(s) of
ingress/egress to other times.  This adjusts our predicted precision,
assuming a single eclipse mapping observation with JWST, to 4
  ppm.

The signal we are trying to measure in eclipse mapping is the
difference in shape between an ingress/egress curve for a planet disk
with uniform brightness and one with a flux gradient.  If the gradient
stretches across the globe, so that one hemisphere is above average
and the other below, then we can simplistically assume that the
maximum signal occurs when the planet is half-occulted by the star (in
reality this depends on the projected angle of the gradient relative
to the stellar limb).\footnote{This signal estimate also requires that the transit
  does not occur exactly edge-on (with an impact parameter of $b=0$), in
contradiction of the edge-on assumption used to create the plots of
predicted planet emission above.  If the eclipse were perfectly edge-on, then a
planet with a projected axial tilt of $0\degr$ would have no
eclipse mapping signal, due to perfect symmetry.  However, the chances
of an exactly edge-on orbit are infinitesimal and so in practice this
is not a significant concern.  This assumption is also not strongly at odds with
our previous analysis, since for a planet to transit at all the
orbital inclination must be such that $b=(a \cos i)/R_s \leq 1$, which
for our hypothetical planet translates to $i \geq 87$\degr.} In
this case, we apply a penalty of $1/2$ to the expected signal, since
only half of the planet is in view.  For planets with an
equator-to-pole flux gradient, the same simplistic assumption would
imply the greatest signal is when $1/4$ of the planet is occulted or
left in-view.  The ingress and egress signals would then average out to the same factor of
$1/2$. The amplitude of the signal we are trying to measure is then roughly
$(1/2) F_{\mathrm{grad}}/F_*$.  (If the gradient is equal to the
eclipse depth, then halfway through ingress/egress the flux would be
$1/2\ F_{\mathrm{ed}}$ different from the flux of a uniformly bright
disk; if there is no gradient the signal is zero.) 

Note that one underlying assumption of our analysis is that
  our predicted JWST observations are at sufficient cadence to resolve the shape of
  the light curve during ingress and egress well enough to retrieve
  the spatial information.  This is a safe assumption for two reasons.
First, \emph{Spitzer} has demonstrated sufficient cadence for mapping a
similarly bright target \citep{deWit2012,Majeau2012}, albeit with
stacked observations, and the time-series imaging mode for NIRCam has
been specifically designed for high-precision and high-cadence
observations of bright sources,\footnote{see {\tt
    https://jwst-docs.stsci.edu/display/JTI/ NIRCam+Time-Series+Imaging}}
a significant improvement over \emph{Spitzer}.  Second, in the
simplified estimate here we are effectively binning down to two points
during ingress and two during egress, since we are only trying to
estimate an amplitude for the flux gradient, rather than its spatial
shape.  As such, the observational cadence of JWST is not a concern here.

We can then take the planet-to-star flux ratios estimated above,
multiply by $0.5 F_{\mathrm{grad}}/F_{\mathrm{ed}}$, 
and divide by our
estimated JWST precision for this type of measurement to determine the
detectability of our predicted flux gradients.  From the 
eclipse depths and flux gradients presented in Figure~\ref{fig:einfo}
we determine that the lower end of the $
F_{\mathrm{grad}}/F_{\mathrm{ed}}$ range for the $\psi=60\degr$
model is not accessible to eclipse mapping with a single observation, the lower end of that
range for the $\psi=90\degr$ model is marginally mappable
(3-$\sigma$), and \emph{all other possible views of all other models could be
mapped with a single JWST observation}. The minimum for the $\psi=30\degr$
model is 6-$\sigma$ and the maximum values for the $\psi=0$\degr,
10\degr, 30\degr, 60\degr, and 90\degr~models are
13-$\sigma$, 12-$\sigma$, 9-$\sigma$, 9-$\sigma$, and 12-$\sigma$, respectively.

The technique of eclipse mapping has only successfully been applied to a single
exoplanet \citep[HD 189733b,][]{deWit2012,Majeau2012} and does not yet
have a standardized method of analysis, making it difficult for us to
predict the precision to which axial tilts could be measured.  Both
\citet{deWit2012} and \citet{Majeau2012} presented multiple possible maps that could be
reconstructed, depending on the particular form of equations used to
describe the planet map.  This motivates work such as Rauscher, Suri,
\& Cowan (in prep), in which we present orthonormal basis sets of maximally
informative light curves and the corresponding ``eigenmaps'' that can
be combined to reconstruct the planet map.  However, our
  ability to accurately use the method of eclipse mapping  is limited by the
  precision with which we know other properties of the system, namely
  the planet's eccentricity, its impact parameter (or alternatively
  its orbital inclination), and the global density of the star (which
  influences the orbit of the planet), as demonstrated by
  \citet{deWit2012}.  It is left to more detailed
future work to precisely quantify the sensitivity of eclipse
mapping measurements to both the amplitudes of flux gradients and
their projected tilts. 

\subsubsection{A less optimistic signal-to-noise estimate}

For the estimates in the sections above, we have optimistically
assumed that JWST will be able to reach photon-limited precision and that our
hypothetical planet orbits a nearby 
bright star.  Here we briefly consider how our estimated
signal-to-noise decreases if we make less generous assumptions.  While
none of these factors can actually be known before JWST is in
operation, or before a warm Jupiter system to study is identified, it
is possible to adjust our expectations for less desirable circumstances.

The hypothetical target star used by \citet{Cowan2015} to estimate
instrument precision, which we use as a basis for our signal-to-noise
estimate, has similar properties to the bright hot Jupiter 
host star HD 189733 ($d=19.45$ pc, $T_{\mathrm{eff}}=5040$ K), which has a $V$
magnitude of 7.7.\footnote{NASA Exoplanet Archive}    The expected
yield from the Transiting Exoplanet Survey Satellite (TESS) includes
about 20 gaseous exoplanets (meaning radii larger than 4 Earth radii)
with periods less than 20 days, orbiting stars with magnitudes as
faint as $I_C \sim 12$, which is roughly the magnitude we would expect for
the HD 189733-like host star used in the calculations above.  So if we
target a warm Jupiter from the TESS survey, we may expect our optimism
about a bright host star to be fair.  If, however, we are limited to
the currently known warm Jupiter population (see the Introduction),
then a host star $V$ magnitude of 10 would make our precision $\sim$3
times worse.  \emph{This would still allow for most of our predicted flux
gradients to be detected at greater than 3-$\sigma$ with a single JWST
observation.}

On the instrument side, while it may be overly optimistic to
  assume that JWST will reach photon-limited precision, the realized
  performance may not be too far off from it.  \citet{Cowan2015}
  showed that their photon-limited precision estimates for HST and
  Spitzer were within a factor of 2 to 3 of the precisions actually
  achieved for exoplanet transits and eclipses.  This is encouraging,
  especially since the NIRCam, NIRISS, and NIRSpec instruments will
  all use the same type of detector as HST.  A factor of 2-3 worse
  precision is the same as our above consideration of a dimmer stellar host,
meaning that \emph{we would still be able to measure most of the
  possible flux gradients to better than 3-$\sigma$}.

We also have reason for some optimism in that, even if the
  precision of JWST is not as spectacular as one might hope, we should
still be able to stack together multiple eclipses to achieve improved
precision.  The original eclipse mapping measurement was achieved by
stacking together seven \emph{Spitzer} eclipses
\citep{deWit2012,Majeau2012}.  Even if the performance of JWST were an
order of magnitude worse that photon-limited precision, stacking
$\sim$4-9 eclipses would still allow us to measure the maximum flux
gradients predicted for each obliquity model (and thus help to
constrain a warm Jupiter's obliquity) at 3-$\sigma$, although $\sim$25 eclipses would
be needed to get these to 5-$\sigma$ detections.

\section{Summary and conclusions} \label{sec:conc}

We have presented a set of three-dimensional atmospheric models for a
hypothetical Jupiter-like planet on a 10-day period orbit
around a Sun-like star, testing possible planetary obliquities ($\psi$) of:
0\degr, 10\degr, 30\degr, 60\degr, and 90\degr. We find
that seasonal variations are negligible for $\psi \leq 10$\degr~and
important for $\psi \geq 30$\degr, to a level that should be
measurable with the \textit{James Webb Space Telescope}.

The circulation pattern of the $\psi \leq 30$\degr~models is
characterized by two high-latitude eastward jets (one in each hemisphere) and an
equator that is hotter than the poles.  For the $\psi=30$\degr~
model the hot region near the equator shifts slightly up and down in
latitude, as a delayed response to the seasonally shifting irradiation
pattern.  This atmospheric response time of $\sim$1/8 \Porb, which is
about an order of magnitude longer than the radiative timescale at the
infrared photosphere, is also seen in the seasonal variations of the
models with $\psi \ge 60$\degr. For those models the temperature
pattern is characterized by one hemisphere being hotter during its
summer and then a more uniform temperature distribution during
equinoxes.

The observable features of each model depends not only on its 
obliquity, but also on the orientation with which the system is
viewed, relative to the seasonal cycle. This results in observational
degeneracies between inherent and coincidental properties; however, by
combining multiple measurements, it may be possible to constrain a
planet's obliquity. From observing the light emitted by the planet
as a function of orbital phase, we can determine:
\begin{itemize}
\item the efficiency of equator-to-pole heat transport, if the phase
  curve is flat and is at a level higher than the global average for
  re-emission of absorbed starlight,
\item the obliquity of the planet, if the phase curve is flat and at a
  level below the global value for re-emitted starlight, or
\item degenerate information about the planet's obliquity, our viewing
  orientation, and the atmospheric seasonal response time, if the
  phase curve shows measurable variation.
\end{itemize}

Eclipse mapping, which resolves the dayside of the planet when it is
eclipsed by the star, can provide more information than orbital phase
curves and will require significantly less telescope time. We present
quantitative predictions for the three main features measured by eclipse
mapping: the total flux emitted from the observed hemisphere, the
gradient in emitted flux across the hemisphere, and the orientation of
that gradient relative to the direction of motion of the stellar
limb.  While none of these parameters can uniquely constrain the
planet's obliquity on its own, their combination can usually only
match the predictions from a single obliquity model. The exceptions
are the $\psi=0\degr$ and 10\degr~models, which 
have very similar values for each of these predicted observables. Thus
\textit{eclipse mapping can either constrain a planet's obliquity to be small
(if $\leq 10\degr$), or can measure it if significantly non-zero ($\ge
30\degr$).}

The phase curve and eclipse mapping methods are also complementarily
informative.  For example, independent of obliquity, the smallest
phase curve variations occur 
for equatorial viewing geometries; but if we observe the system from
this orientation, the projected axial tilt measured by eclipse mapping
will be equal to the true obliquity of the planet.
In this way, the multiple parameters that
can be observed with eclipse mapping could be combined with phase
curve measurements to provide even better constraints on the planet's
obliquity. If we first obtain the (observationally cheaper) eclipse
map measurement, we could use the information contained there to
identify which orbital phases might be the most informative (providing
information about the atmospheric response time and/or efficiency of
meridional heat transport) and sidestep the need for continuous
observation \citep{Krick2016}.
 
Finally, we compare our various predicted signals to the sensitivity
of current and future instruments.  We find that Hubble observes at
wavelengths too short to be useful for observing this
hypothetical planet; however, the James Webb Space Telescope will be an
amazing instrument for this type work.  If we assume
  photon-limited precision and a host star as bright as HD~189733,
  then \textit{JWST should be able to detect the
secondary eclipse of this hypothetical warm Jupiter at $\sim$80-$\sigma$ in a
single observation}, which is sensitive enough to eclipse-map the
planet, measuring the hemispheric flux gradient and its orientation. 

We have determined that the expected non-zero obliquities of warm
Jupiters should both influence the observational characterization of these
planets, and also be measurable with JWST. Future work will need to
determine the observational degeneracies between the diversity of possible
values we expect for the obliquities, eccentricities
\citep{Gaidos2004,Kataria2013}, and rotation rates \citep{Showman2015}
of warm Jupiters. Since the obliquities (along with eccentricities and
rotation rates) are presumably markers of the formation, evolution,
and tidal processes that influence these planets, this adds an extra
dimension of study for warm Jupiters, beyond the lessons we can learn
from their bulk compositions \citep[e.g.][]{Thorngren2016}.

\acknowledgments

I thank Miles for significantly delaying this publication (by being
born and so completely disrupting my life). I wouldn't have it
any other way.

I thank the anonymous referee for constructive comments that helped to
improve and clarify this manuscript.

This research was supported by NASA Astrophysics Theory Program grant
NNX17AG25G and made use of the NASA Exoplanet Archive, which is
operated by the California Institute of Technology, under contract
with the National Aeronautics and Space Administration under the
Exoplanet Exploration Program.

\bibliography{biblio.bib}

\end{document}